\documentclass[a4paper,11pt]{article}
\usepackage{array,amssymb,amsmath,amsfonts,commath,graphicx}
\usepackage{natbib}
\usepackage[margin=2.5cm]{geometry}

\begin{document}
\title{Torsion of a cylinder of partially molten rock with a spherical
  inclusion: theory and simulation}

\author{Laura Alisic\thanks{Department of Earth Sciences, University
    of Cambridge, Cambridge, United Kingdom (now at Jet
    Propulsion Laboratory, California Institute of Technology, USA).}
  \and Sander Rhebergen\thanks{Department of Applied Mathematics,
    University of Waterloo, Canada.}
  \and John F.~Rudge\thanks{Department of Earth Sciences, University
    of Cambridge, United Kingdom.}
  \and Richard F.~Katz\thanks{Department of Earth Sciences,
    University of Oxford, United Kingdom.}
  \and Garth N.~Wells\thanks{Department of Engineering,
    University of Cambridge, Cambridge, United Kingdom.} }

\date{}

\maketitle

\begin{abstract}
The processes that are involved in migration and extraction of melt
from the mantle are not yet fully understood. Gaining a better
understanding of material properties of partially molten rock could
help shed light on the behavior of melt on larger scales in the
mantle. In this study, we simulate three-dimensional torsional
deformation of a partially molten rock that contains a rigid,
spherical inclusion. We compare the computed porosity patterns to
those found in recent laboratory experiments. The laboratory
experiments show emergence of melt-rich bands throughout the rock
sample, and pressure shadows around the inclusion. The numerical model
displays similar melt-rich bands only for a small
bulk-to-shear-viscosity ratio (five or less). The results are
consistent with earlier two-dimensional numerical simulations;
however, we show that it is easier to form melt-rich bands in three
dimensions compared to two. The addition of strain-rate dependence of
the viscosity causes a distinct change in the shape of pressure
shadows around the inclusion. This change in shape presents an
opportunity for experimentalists to identify the strain-rate
dependence and therefore the dominant deformation mechanism in torsion
experiments with inclusions.
\end{abstract}

\section{Introduction}

The transport of melt in the mantle plays an important role in the
dynamics and chemical evolution of both the mantle and the
crust. Although the equations that describe the conservation of mass,
momentum, and energy of partially molten rock are well established
\citep{McKenzie:1984, Bercovici:2003}, the appropriate constitutive
relations remain uncertain. This means that the dynamics of melt
segregation and transport present significant unanswered questions.

One means of addressing questions on the dynamics of melt segregation
and transport is by comparison of simulations with laboratory
experiments on partially molten rocks subjected to forced
deformation. A recent experimental study with significant potential in
this regard is reported by \citet{Qi:2013}.  Following on the
torsional deformation experiments of \citet{King:2010},
\citet{Qi:2013} modified the basic experiment by including rigid,
spherical beads within the partially molten rock that is undergoing
deformation.  They find that pressure shadows around the bead are
expressed as variations in melt fraction there.  Furthermore, they
find that melt-rich bands, also observed in experiments without beads
\citep[e.g.][]{Holtzman:2003}, emerge and tend to connect with the
large-porosity lobes of the pressure shadow. Previous analysis of
pressure shadows \citep{McKenzie:2000, Rudge:2014} and their
interaction with banding instabilities \citep{Alisic:2014} suggests
that the observed relationship between these two modes of compaction
could constrain the bulk viscosity of the two-phase system.

In working towards a better understanding of compaction in a two-phase
system, we pose the following questions about the viscosity of the
two-phase system that remain unresolved.  What is the ratio of the
bulk viscosity to the shear viscosity at small reference porosity
\citep{Simpson:2010}?  How do the bulk and shear viscosities vary with
porosity \citep[e.g.][]{Kelemen:1997, Mei:2002, Takei:2009}?  Is the
rheology non-Newtonian and, if so, does this help to explain the
patterns observed in experiments?  And, more broadly, is a solely
viscous rheology sufficient to capture the dynamics?  These are
long-term questions that we address.  However, we find that on the
basis of the comparison between experiments and theory considered
here, we cannot answer these questions definitively, and we present a
discussion of this shortcoming.

In an earlier paper, we developed two-dimensional models of two-phase
flow around a cylindrical inclusion to study the same experimental
system \citep{Alisic:2014}. Here we build on those results by
expanding the numerical simulations to three dimensions. This allows
us to capture the three-dimensional scaling of compaction around a
sphere, which differs from the two-dimensional scaling around a
cylinder \citep{Rudge:2014}. Moreover, the simulations presented here
provide a more realistic comparison to the results of laboratory
experiments \citep{Qi:2013}. These new simulations with $\sim 7 \times
10^{6}$ degrees of freedom would be impossible without an advanced,
new preconditioning method for the equations of magma dynamics that
has been recently developed \citep{Rhebergen:2015}.

We begin this manuscript with a description of the equations governing
deformation and compaction of partially molten rock, after which we
summarize the domain geometry, boundary conditions, and discretization
used in the numerical simulations. Analytical solutions for certain
limiting cases are provided in Appendix \ref{sec:analysis}; we use
these to benchmark the simulation code. The first set of results in
Section \ref{sec:uniform} pertains to simulations with a uniform
initial porosity that allow us to focus on compaction around a
spherical inclusion in three dimensions, with Newtonian and
non-Newtonian rheology. The simulations in Section
\ref{sec:non-uniform} focus on problems with a random initial porosity
field, where we investigate the interaction between pressure shadows
around the inclusion and melt-rich bands developing throughout the
domain. The results are followed by a discussion in Section
\ref{sec:discussion}, after which conclusions are drawn.

\section{The model}

\subsection{Governing equations}

The compaction of partially molten rock and the transport of melt can
be described by governing equations for two-phase flow, formulated
here following \citet{McKenzie:1984}. In dimensionless form (see
Appendix \ref{sec:nondim} for the nondimensionalization):
\begin{eqnarray}
  \frac{\partial \phi}{\partial t} -
  \nabla \cdot \del{1 - \phi} \mathbf{u}_{s} &= 0,
  \label{eq:sys1_nondim}
  \\
  - \nabla \cdot \mathbf{u}_s +
  \nabla \cdot \left( \frac{D^2}{R + \frac{4}{3}}
  K_{\phi} \nabla p_{f} \right) &= 0,
  \label{eq:sys2_nondim}
  \\
  - \nabla \cdot \mathbf{u}_{s} -
  \left( R \zeta \right)^{-1} p_c &= 0,
  \label{eq:sys3_nondim}
  \\
  - \nabla \cdot \Bar{\boldsymbol{\tau}}
  + \nabla p_f + \nabla p_c &= \mathbf{0},
  \label{eq:sys4_nondim}
\end{eqnarray}
where $t$ denotes time, $\phi$ is porosity, $\mathbf{u}_{s}$ is the
solid (matrix) velocity, $p_f$ and $p_c$ are the magma and compaction
pressure, respectively, and $\Bar{\boldsymbol{\tau}}$ is the
deviatoric stress in the solid. Constitutive properties, discussed
further below, appear as $K_{\phi}$ for the permeability and $\zeta$
for the bulk viscosity.  The bulk-to-shear-viscosity ratio in the
reference state is defined as $R = \zeta_{\mathrm{ref}} /
\eta_{\mathrm{ref}}$, where $\zeta_{\mathrm{ref}}$ is a reference bulk
viscosity and $\eta_{\mathrm{ref}}$ is a reference shear viscosity for
the two-phase mixture.  Finally, $D = \delta / H$ where $\delta$ is
the compaction length and $H$ is the height of the domain. The
compaction length is given by~\citep{McKenzie:1984}:
\begin{equation}
  \delta
  = \sqrt{\frac{\del{R + \frac{4}{3}} \eta_{\mathrm{ref}}
  K_{\mathrm{ref}}}{\mu_{f}}},
  \label{eq:compaction_length}
\end{equation}
where $K_{\mathrm{ref}}$ is the permeability in the reference state,
and $\mu_f$ is the magma viscosity. In this study we assume a
compaction length that is much larger than the domain size ($D =
100$).  The deviatoric stress tensor $\Bar{\boldsymbol{\tau}}$ is
\begin{equation}
  \Bar{\boldsymbol{\tau}}
  = \eta \underbrace{\del{\nabla \mathbf{u}_{s}
      + \del{\nabla \mathbf{u}_{s}}^T
      - \frac{2}{3} \del{\nabla \cdot \mathbf{u}_{s}}
      \mathbf{I}}}_{2 \dot{\boldsymbol{e}}},
\end{equation}
where $\eta$ is the shear viscosity and $\dot{\boldsymbol{e}}$ is the
deviatoric strain-rate tensor.

The above model assumes that no melting or solidification takes place,
buoyancy forces are negligible, and that the fluid and solid phases
have densities that are constant (but different from each other);
these assumptions are appropriate for the motivating laboratory
experiments \citep[e.g.][]{Holtzman:2003}, though a model for the
Earth's mantle clearly must be more general. The unknown fields in the
model are $\phi$, $\mathbf{u}_{s}$, $p_f$, and $p_c$, which must
satisfy equations~\eqref{eq:sys1_nondim}--\eqref{eq:sys4_nondim},
subject to the boundary conditions described below.

\subsection{Rheology and permeability}

Closure conditions for the governing equations are prescribed as
\begin{equation}
  \begin{gathered}
    K_{\phi} = \left( \frac{\phi}{\phi_0} \right)^n,
    \quad
    \eta = \left(2 \dot{{\varepsilon}}\right)^{-q}
    \mathrm{e}^{-\alpha (\phi - \phi_0)},
    \quad
    \zeta = \underbrace{\left(2\dot{{\varepsilon}} \right)^{-q}
    \mathrm{e}^{-\alpha (\phi - \phi_0)}}_{\eta}
    \left( \frac{\phi}{\phi_0} \right)^{-m},
  \end{gathered}
\end{equation}
where $n$ and $m$ depend on the melt geometry considered. We take $n =
2$ and $m = 1$, assuming a tubular melt geometry. In these
definitions, $\phi_0$ is the reference porosity, $\alpha$ is a
constant representing the sensitivity of matrix shear viscosity to
porosity, $\dot{{\varepsilon}}$ is the second invariant of the
deviatoric strain-rate tensor,
\begin{equation}
  \dot{{\varepsilon}} = \left(\frac{1}{2} \dot{\boldsymbol{e}}:
  \dot{\boldsymbol{e}} \right)^{1/2},
\end{equation}
and $q$ is related to the power-law exponent $\mathfrak{n}$ by
\begin{equation}
  q = 1 - \frac{1}{\mathfrak{n}}.
\end{equation}
A power-law exponent $\mathfrak{n} = 1$ gives the limit of Newtonian
rheology.

In this study we focus on the effects of the porosity sensitivity
$\alpha$, the reference bulk-to-shear-viscosity ratio $R$, and the
power-law exponent $\mathfrak{n}$ on compaction patterns around and
away from an inclusion. Laboratory experiments indicate that $\alpha$
is around $26$ for a Newtonian rheology (diffusion creep) and around
$31/\mathfrak{n}$ for dislocation creep \citep{Kelemen:1997,
  Mei:2002}. Several previous modelling studies have used a value
$\alpha = 28$ (e.g.~\citet{Alisic:2014, Katz:2006}). The
bulk-to-shear-viscosity ratio, however, is poorly
constrained. Theoretical and experimental studies place $R$ between
order one, independent of the reference porosity \citep{Takei:2009},
and $\sim 20$ for a reference porosity $\phi_0 = 0.05$ on the basis of
the expected relation $R \propto \phi^{-1}$
\citep{Bercovici:2003,Simpson:2010}. In the simulations presented
here, we vary $\alpha$ between zero and 50, and $R$ between $5/3$ and
20. In simulations where we study the effect of strain-rate dependence
of the shear and bulk viscosities, the power-law exponent
$\mathfrak{n}$ has values between 1 and 6.  The deformation mechanism
of diffusion creep corresponds to $\mathfrak{n} = 1$, resulting in
Newtonian viscosities. A larger exponent of around 3 to 4 is relevant
for dislocation creep. \citet{Katz:2006} showed that increasing the
power-law exponent is one possible mechanism for reproducing the
shallow angle of melt-rich bands as observed in laboratory
experiments. We therefore include simulations with $\mathfrak{n}$ up
to six.

\subsection{Domain of interest and boundary and initial conditions}
\label{sec:domain}

\begin{figure}
  \center\includegraphics[width=0.7\textwidth]
  {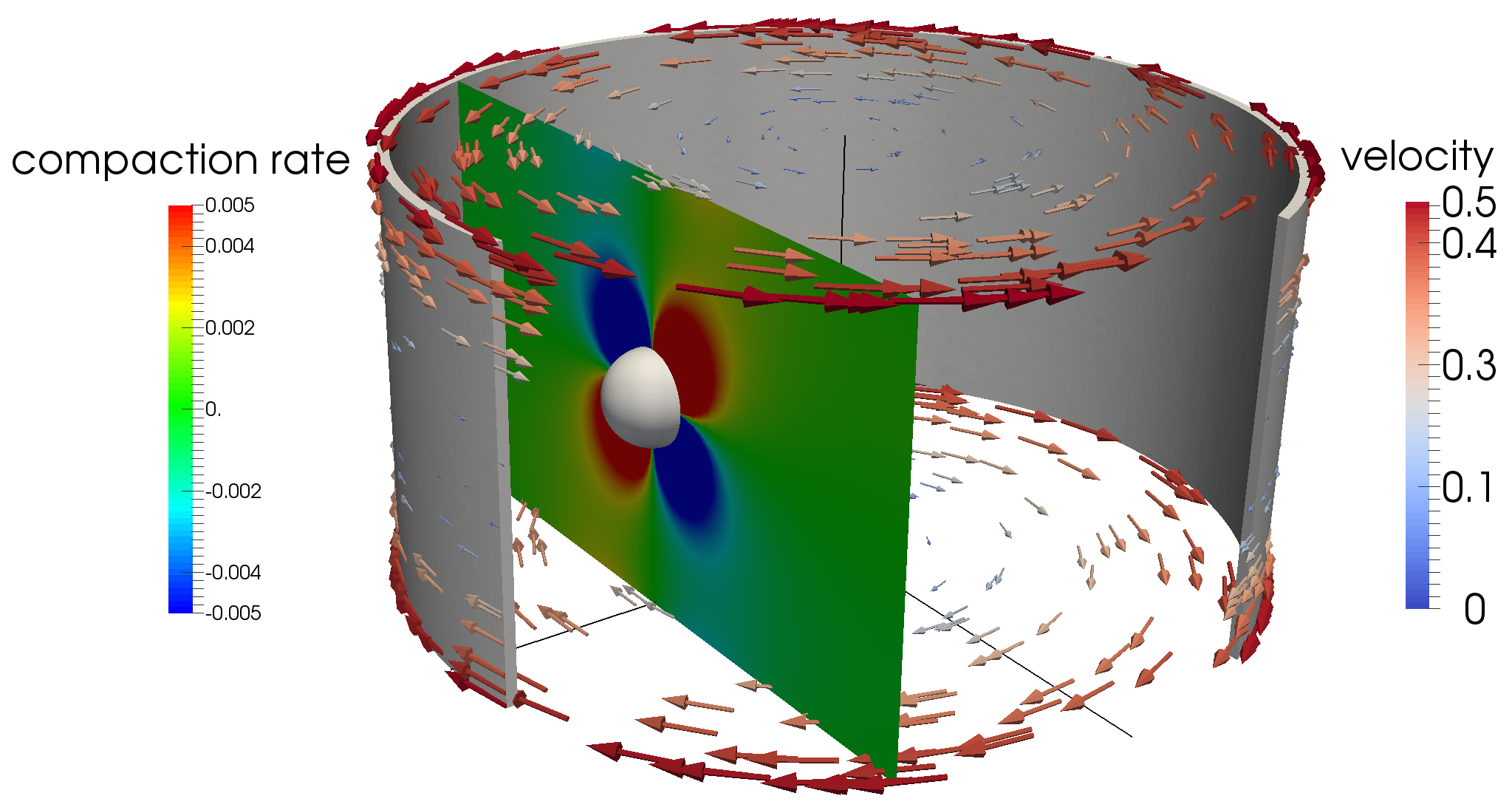}
  \caption{The geometry of the model domain: a cylinder of height~$H =
    1$ and radius one (only a thin cut-out is shown in light gray),
    with a rigid spherical inclusion of radius~$0.1$. On a
    two-dimensional slice through the cylinder and inclusion at~$x =
    \frac{1}{2}$, the instantaneous compaction rate ($\nabla \cdot
    \mathbf{u}_{s}$) is plotted for a simulation with
    bulk-to-shear-viscosity ratio~$R = 20$, and power-law
    exponent~$\mathfrak{n} = 1$ at time~$t = 0$. The arrows at the
    top, bottom and side of the cylinder indicate the prescribed solid
    velocity on the cylinder boundaries.}
\label{fig:3Dgeometry}
\end{figure}

We compute solutions to
equations~\eqref{eq:sys1_nondim}--\eqref{eq:sys4_nondim} in a
cylindrical domain $\Omega \subset \mathbb{R}^{3}$ of height $H = 1$
and radius $1$, where $x^2 + y^2 \leq 1$ and $0 \leq z \leq 1$. A
rigid spherical inclusion is centered at $\mathbf{r}_0 = (1/2, 0,
1/2)$ and has a radius of $0.1$, as shown in Figure
\ref{fig:3Dgeometry}.
The inclusion is modeled as a spherical hole in the domain that cannot
deform. The boundary conditions are defined on the boundary
$\partial\Omega$ as:
\begin{align}
  K_{\phi} \nabla p_{f} \cdot \mathbf{n}
  &= 0 \ \text{on} \ \partial\Omega,
  \label{eq:imperm}
  \\
  \mathbf{u}_{s} &= \mathbf{w} \ \text{on} \ \partial\Omega,
  \label{eq:dirbc}
\end{align}
where the boundaries are taken to be impermeable (equation
\eqref{eq:imperm}), and $\mathbf{w}$ is a prescribed solid
velocity. The cylinder is placed under torsion. This torsion is
enforced by Dirichlet boundary conditions of the form \eqref{eq:dirbc}
on the top and bottom ($z = 0$ and $z = 1$), and side boundaries of
the cylinder ($x^{2} + y^{2} = 1$), such that on the outside of the
cylinder $\mathbf{w} = \mathbf{u}_{\textrm{cyl}}$:
\begin{equation}
  \mathbf{u}_{\textrm{cyl}} = \del{-y \del{z - \tfrac{1}{2}}, \,
  x \del{z - \tfrac{1}{2}}, \,
  0}.
\end{equation}
Formally, the boundary conditions on the rigid inclusion are
conditions of no net force and no net torque (see
\citet[Appendix~B]{Alisic:2014}). Here, to simplify the construction
of the numerical model, we instead apply a Dirichlet boundary
condition of the form \eqref{eq:dirbc} that approximates the zero net
force and torque conditions. In a uniform medium, a rigid sphere
placed at the mid-plane of a torsion field should not translate, but
should rotate with an angular velocity equal to half the vorticity of
the imposed torsion field (see Section \ref{sec:instcomp}). Thus we
use a Dirichlet condition on the boundary of the inclusion with
$\mathbf{w} = \mathbf{u}_{\textrm{sphere}}$,
\begin{equation}
  \mathbf{u}_{\textrm{sphere}} = \boldsymbol{\Omega} \times
  \left(\mathbf{r} - \mathbf{r}_{0} \right),
  \label{eq:rotbc}
\end{equation}
where $\mathbf{r}=(x, y, z)$ is the position vector, and
$\mathbf{r}_{0}$ is the center of the sphere. The angular velocity
$\boldsymbol{\Omega}$ is given by
\begin{equation}
  \boldsymbol{\Omega}
  = \frac{1}{2} \left. \nabla \times
  \mathbf{u}_{\textrm{cyl}} \right\vert_{\mathbf{r} = \mathbf{r}_{0}}
  = \del{-\tfrac{1}{4}, 0, 0}.
\end{equation}
In our Cartesian coordinate system and for the position of the
inclusion, \eqref{eq:rotbc} can thus be written
\begin{equation}
  \mathbf{u}_{\textrm{sphere}} = \del{0, \, \tfrac{1}{4} \del{z
    - \tfrac{1}{2}}, \,
    - \tfrac{1}{4} y}.
\end{equation}
The placement of the inclusion at $x = 1/2$ results in a local strain
at the center of the inclusion equal to half the total, outer-radius
model strain, which scales to half the model time in a simulation.

We choose either a constant initial porosity field with $\phi_0 =
0.05$, or a random initial porosity with uniformly distributed values
in the range $\phi_0 \pm 5 \times 10^{-3}$.  For the simulations with
a randomly perturbed initial porosity field, we produced one initial
field and reused this for all simulations. This initial field is
created by first generating a random field on a uniform mesh that has
a slightly larger grid size than the largest elements in the cylinder
mesh; then this is interpolated onto the cylindrical mesh containing
the spherical hole and variable grid size. This approach ensures that
the random perturbations are sufficiently resolved by the mesh used in
simulations and that the length scale of the perturbations does not
vary with element size.

Throughout this paper, we present simulation results on a
two-dimensional slice through the inclusion at $x = 1/2$, as shown in
Figure \ref{fig:3Dgeometry}. In this figure, the instantaneous
compaction rate at time $t = 0$ for a simulation with
bulk-to-shear-viscosity ratio $R = 20$ is shown on the slice. The
initial compaction rate is independent of the porosity exponent
$\alpha$ for uniform porosity initial conditions. Pressure gradients
caused by flow past the spherical inclusion induce two compacting
lobes and two dilating lobes around the inclusion. This behavior was
described in detail by \citet{McKenzie:2000, Alisic:2014, Rudge:2014}
and is discussed further in Section \ref{sec:instcomp}.

\subsection{Discretization}

The problem described in Section \ref{sec:domain} is solved by a
finite element method on a mesh of tetrahedral cells consisting of
approximately 50 cells in the vertical dimension.  The mesh is refined
around the inclusion. The smallest cell size is $\sim 3 \times
10^{-3}$ near the inclusion, and the largest cell size is $\sim 7
\times 10^{-2}$ away from it.

There are two main time stepping approaches to solving the two-phase
flow equations~\eqref{eq:sys1_nondim}--\eqref{eq:sys4_nondim}, namely,
as a fully coupled system~\citep{Katz:2007} or by decoupling the
porosity evolution equation~\eqref{eq:sys1_nondim} from the compaction
equations~\eqref{eq:sys2_nondim}--\eqref{eq:sys4_nondim}
\citep{Katz:2013}. We follow the second approach. At each time step,
equations \eqref{eq:sys2_nondim}--\eqref{eq:sys4_nondim} are solved to
find the solid velocity, fluid pressure and compaction pressure, given
the porosity and viscosities from the previous iteration. The porosity
is then updated by solving~\eqref{eq:sys1_nondim}. To ensure a good
approximation of the coupling, we iterate this process. Furthermore,
if a non-Newtonian rheology is used, within each iteration a new
strain rate is computed from the solid velocity and the viscosities
are updated accordingly.

The compaction system~\eqref{eq:sys2_nondim}--\eqref{eq:sys4_nondim}
is discretized with a continuous Galerkin finite element method using
Taylor--Hood type elements (piecewise quadratic polynomial
approximation for the solid velocity and piecewise linear polynomial
approximation for the fluid and compaction pressures, see
\citet{Rhebergen:2014}). The system of linear equations resulting from
this discretization is solved using Bi-CGSTAB in combination with the
block-preconditioners developed in~\citet{Rhebergen:2015}.

The porosity evolution equation~\eqref{eq:sys1_nondim} is discretized
in space by a discontinuous Galerkin finite element method using a
linear polynomial approximation.  A Crank--Nicolson time stepping
scheme is used to discretize in time (but using only the most recently
computed velocity). To stabilize the simulation, a
porosity-gradient-dependent artificial diffusion is added to the
porosity evolution equation~\eqref{eq:sys1_nondim} of the form
$\epsilon \nabla \cdot (|\nabla \phi |^3 \nabla \phi)$, with
$\epsilon=0.1$. To solve the resulting discrete system we use
restarted GMRES preconditioned by algebraic multigrid.  Simulations
are terminated when the porosity becomes smaller than zero or larger
than unity.

Instead of solving~\eqref{eq:sys2_nondim}--\eqref{eq:sys4_nondim}, it
is possible to eliminate the compaction pressure by
substituting~\eqref{eq:sys3_nondim} into~\eqref{eq:sys4_nondim}. The
reduced system has fewer unknowns, but solving it is numerically less
robust and less efficient than solving the expanded
system~\eqref{eq:sys2_nondim}--\eqref{eq:sys4_nondim}. We refer
to~\citet{Rhebergen:2015} for more details.

Our simulation code is developed within the finite element software
framework FEniCS/DOLFIN \citep{fenics:book,logg:2010}, in conjunction
with the PETSc linear algebra and solver library
\citep{petsc-web-page,petsc-user-ref}.

\section{Results}
\label{sec:results}

We group our results into two categories.  In the first, the porosity
is initially uniform. This means that the initial growth rate of the
melt-banding instability is zero, and hence that changes in porosity
are initially due solely to the presence of the inclusion. We consider
the sensitivity of the compaction pattern to problem parameters,
including the stress-dependence of the viscosity.  The second category
uses an initial condition with a random porosity
perturbation. Melt-rich bands can potentially develop from the outset
in this class of simulations. We explore in detail how and when such
melt-rich bands develop, and how they interact with pressure shadows
around the inclusion.

\subsection{Uniform initial porosity}
\label{sec:uniform}

\begin{figure}
  \center\includegraphics[width=0.9\textwidth]
  {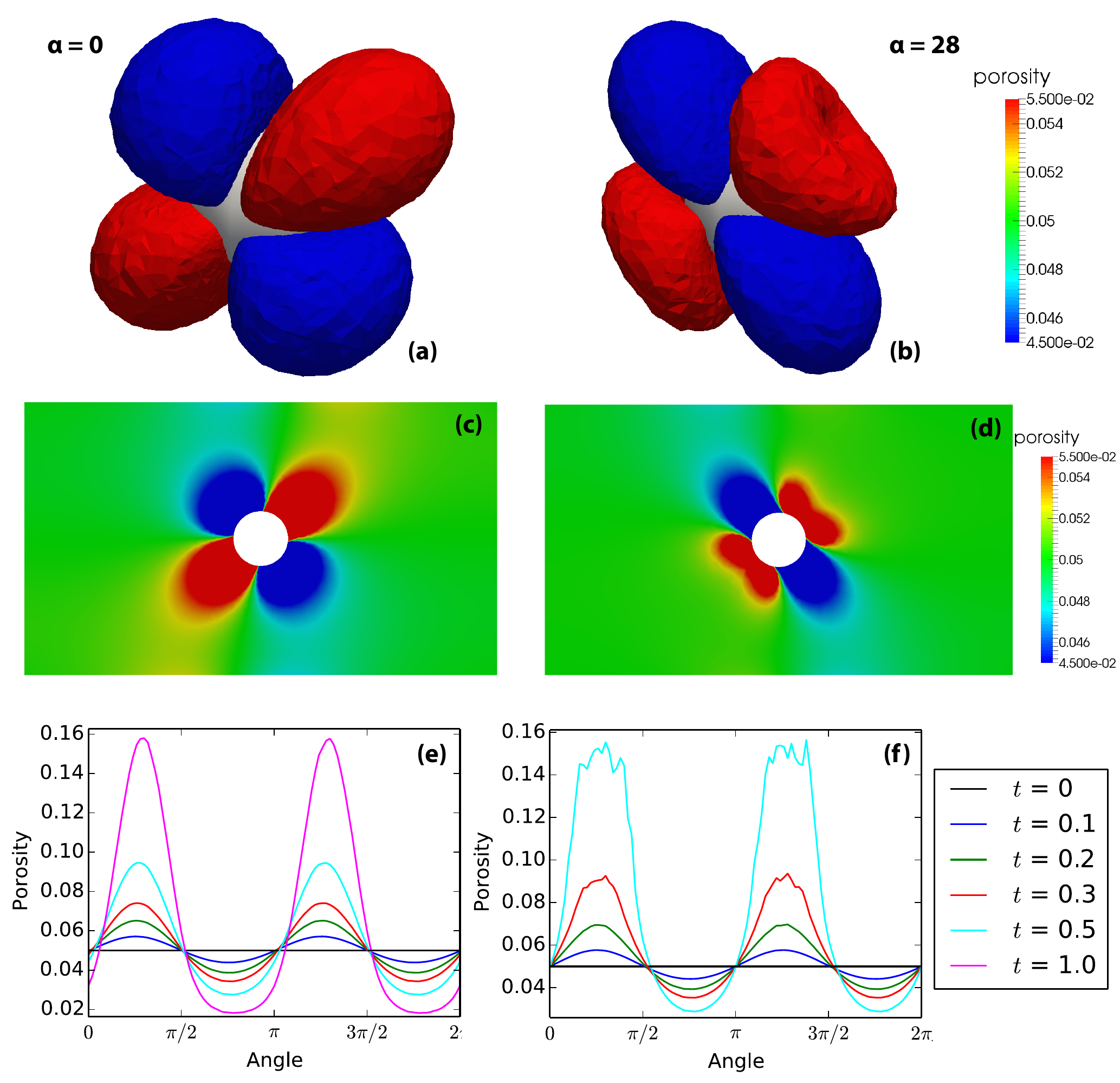}
  \caption{Results for simulations with a uniform initial porosity
    field and Newtonian shear viscosity ($\mathfrak{n} = 1$). The
    local strain at the center of the inclusion corresponds to one
    half of the reported model time.  \textbf{(a)} Three-dimensional
    view of the porosity field for a simulation with porosity exponent
    $\alpha = 0$, bulk-to-shear-viscosity ratio $R = 5$, at time $t =
    0.5$ corresponding to a local strain of 0.25 at the center of the
    inclusion. The pressure shadows around the inclusion are shown as
    porosity contours of 0.045 in blue and 0.055 in red.  \textbf{(b)}
    Three-dimensional view of the porosity field for a simulation with
    $\alpha = 28$ and $R = 5$, at time $t = 0.5$, with porosity
    contours at $0.045$ and $0.055$.  \textbf{(c)} Slice through the
    porosity field at $x = \frac{1}{2}$, for the simulation with
    $\alpha = 0$, $R = 5$, at $t = 0.5$.  \textbf{(d)} Slice through
    the porosity field, for the simulation with $\alpha = 28$, $R =
    5$, at $t = 0.5$.  \textbf{(e)} Radial integrals over porosity,
    for the simulation with $\alpha = 0$, $R = 5$, at various
    times~$t$.  \textbf{(f)} Radial integrals over porosity, for the
    simulation with $\alpha = 28$, $R = 5$, at various times~$t$.}
  \label{fig:uniform}
\end{figure}

\paragraph{Newtonian viscosity} We investigate the effect of the
porosity exponent $\alpha$ and bulk-to-shear-viscosity ratio $R$ on
the porosity evolution in time-dependent simulations with a uniform
initial porosity of $\phi_0 = 0.05$ and Newtonian viscosity. When
$\alpha = 0$ and $\mathfrak{n} = 1$, the shear viscosity is constant
and uniform. In this case, the pressure shadows around the inclusion
that are identified by perturbations in the porosity field rotate and
advect with the matrix, with the top moving to the right and the
bottom to the left, as shown in Figure \ref{fig:uniform}a and~c (note
that all cross-sections presented in this paper are oriented to have
the same direction of shear). In contrast, in a simulation with
$\alpha = 28$ and $\mathfrak{n} = 1$, shown in Figure
\ref{fig:uniform}b and~d, the pressure shadows change shape in the
opposite direction over time, following the orientation of expected
bands in an inhomogeneous model \citep[e.g.][]{Spiegelman:2003}.

\begin{figure}
  \center\includegraphics[width=0.9\textwidth]
  {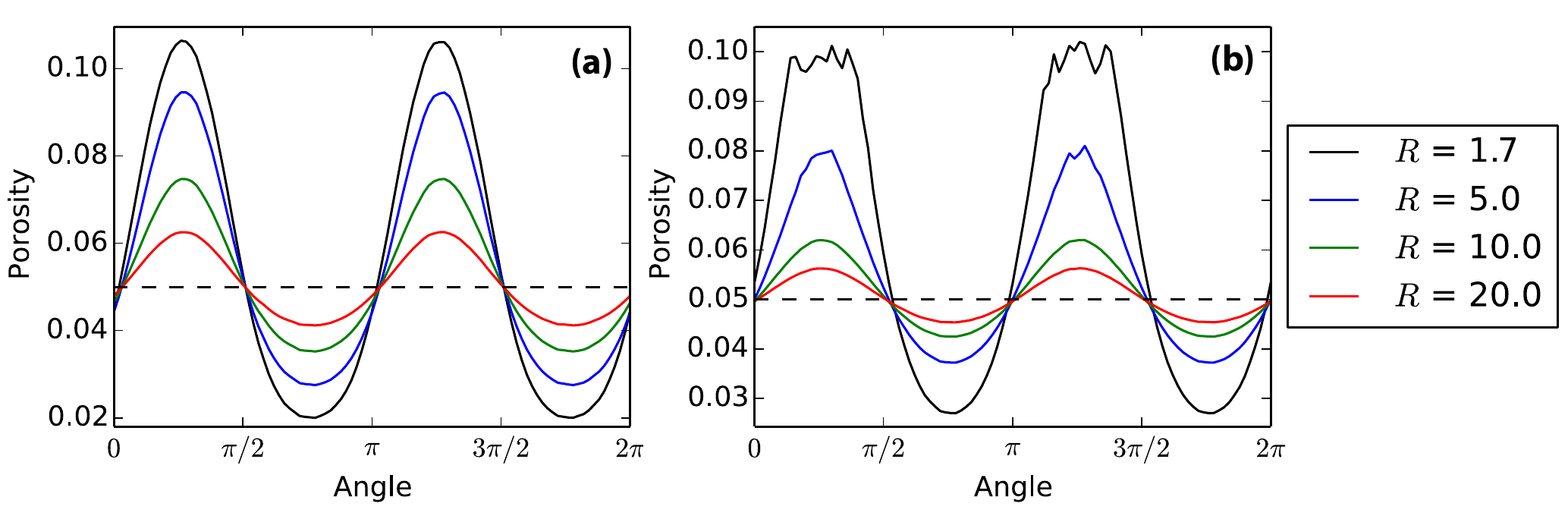}
  \caption{Radial integrals over porosity for simulations with uniform
    initial porosity field and $\mathfrak{n}$ = 1, with various values
    of bulk-to-shear-viscosity ratio~$R$.  \textbf{(a)} Simulations
    with $\alpha = 0$ at time $t = 0.5$, for various values of
    $R$. \textbf{(b)} Simulations with $\alpha = 28$ at time $t =
    0.25$ or a local strain of $0.125$ at the center of the inclusion,
    for various values of~$R$. }
  \label{fig:uniform_R}
\end{figure}

To study the behavior of pressure shadows in more detail, we compute
integrals of porosity on the two-dimensional slice through the
inclusion. Integration is from the local radius of the edge of the
inclusion $r = a$ to one inclusion radius outward at $r = 2a$, for a
series of azimuths between $0$ and $2 \pi$ around the circular
cross-section of the inclusion \citep{Qi:2013, Alisic:2014}
\begin{equation}
  \frac{1}{a} \int^{2a}_{a} \phi \dif r.
\end{equation}
Such radial integrals of porosity around the inclusion help expose the
effect of $\alpha$ on the time evolution of pressure shadows. For the
$\alpha = 0$ simulation, the peaks become sharper over time, and the
troughs become wider (see Figure \ref{fig:uniform}e). The opposite
happens for the $\alpha = 28$ model (see Figure \ref{fig:uniform}f),
with widening peaks. These differences are more pronounced for smaller
bulk-to-shear-viscosity ratios $R$, as seen in Figure
\ref{fig:uniform_R}. These results are consistent with the
two-dimensional results presented in \citet{Alisic:2014}.

\begin{figure}
  \center\includegraphics[width=0.9\textwidth]
  {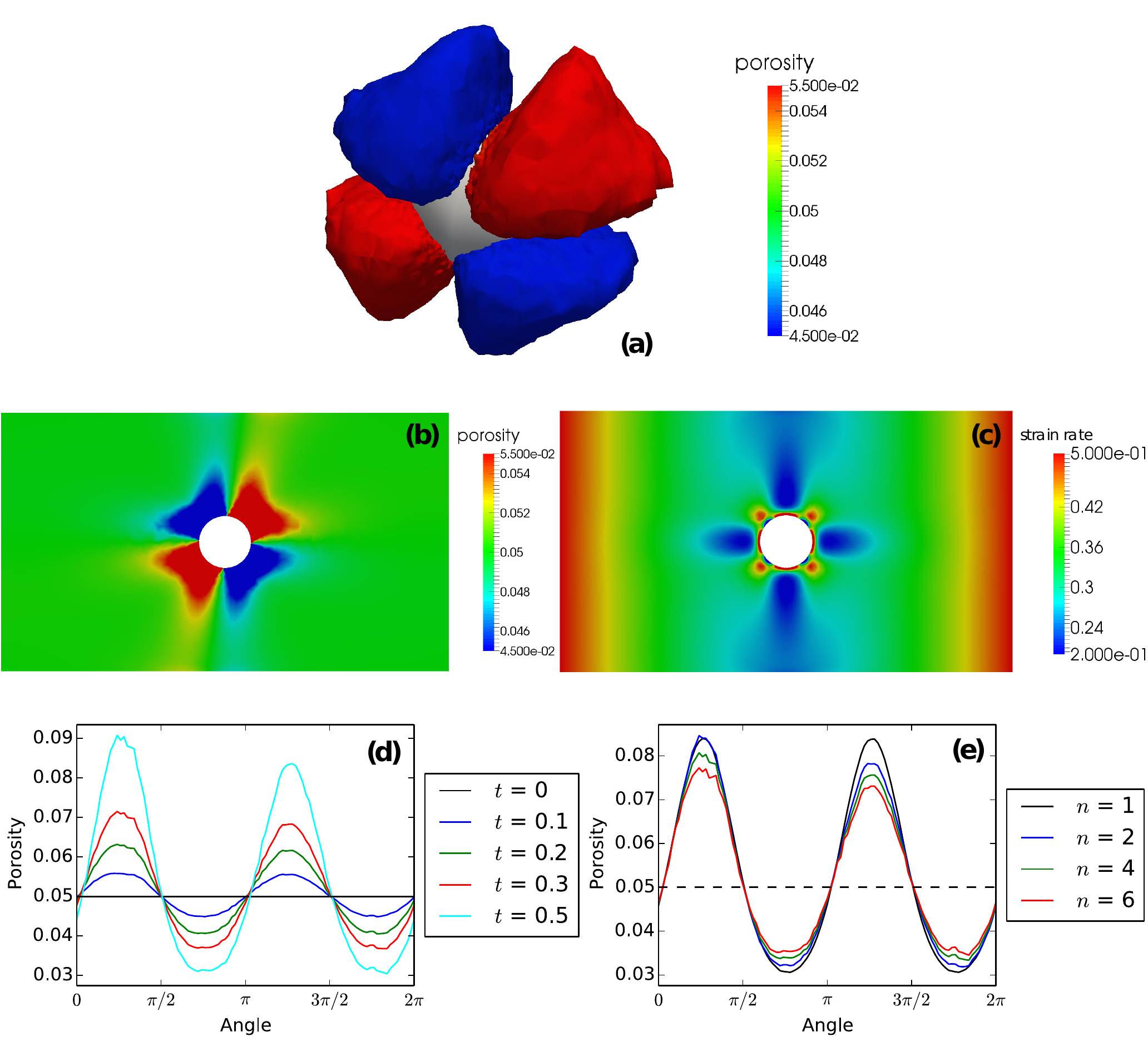}
  \caption{Results for simulations with a uniform initial porosity
    field and a non-Newtonian shear viscosity. The local strain at the
    center of the inclusion corresponds to one half of the reported
    model time. \textbf{(a)} Three-dimensional view of the porosity
    field for a simulation with porosity exponent $\alpha = 0$,
    bulk-to-shear-viscosity ratio $R = 5$, and power-law exponent
    $\mathfrak{n} = 4$ at time $t = 0.5$ corresponding to a local
    strain of $0.25$ at the center of the inclusion. The pressure
    shadows around the inclusion are shown as porosity contours of
    $0.045$ in blue and $0.055$ in red.  \textbf{(b)} Slice through
    the porosity field at $x = \frac{1}{2}$, for the same simulation
    at $t = 0.5$. \textbf{(c)} Second invariant of the strain-rate
    field at $t = 0$.  \textbf{(d)} Radial integrals over porosity,
    for the same simulation at various times. \textbf{(e)} Simulations
    with $\alpha = 0$ and $R = 5$ at time $t = 0.4$, for various
    values of~$\mathfrak{n}$.}
\label{fig:uniform_n4}
\end{figure}

\paragraph{Non-Newtonian viscosity} We introduce a non-Newtonian,
power-law rheology in time-dependent simulations with uniform initial
porosity. In these simulations, the power-law exponent $\mathfrak{n}$
is larger than one. The geometry of pressure shadows around the
inclusion is affected by this strain-rate dependence, and `spokes'
form on either side of each pressure shadow quadrant in a simulation
with $\alpha = 0$, as shown in Figure \ref{fig:uniform_n4}a-b. This
pattern is similar to the shape of pressure anomalies in non-Newtonian
materials under simple shear found by \citet{Tenczer:2001}. The second
invariant of the strain-rate, shown in Figure \ref{fig:uniform_n4}c
and which controls the viscosity variations, exhibits a complex
pattern around the inclusion, without significant temporal variation
throughout the simulation time.

An increase in the power-law exponent $\mathfrak{n}$ results in more
pronounced spokes in the pressure shadows. However, these spokes
mostly develop further than one inclusion radius away from the edge of
the inclusion, and therefore the spoke shape is not reflected in the
radial integrals (see Figure \ref{fig:uniform_n4}d-e). Figure
\ref{fig:uniform_n4}e further indicates that there is a decrease in
amplitude of the peaks and troughs of the radial integrals for an
increase in~$\mathfrak{n}$. This implies that the strain-rate
dependence of the viscosity does not enhance porosity growth rates.

Increasing the porosity exponent $\alpha$ up to $28$ (not shown here)
does not result in a significant change of geometry of the pressure
shadows in simulations with a total strain up to~$0.2$, indicating
that the strain-rate dependence of the rheology is dominant over the
porosity dependence at low strains. It is to be expected that at
larger strains, when porosity anomalies have developed larger
amplitudes, the porosity dependence becomes more significant. This
could then lead to larger differences in geometry for $\alpha = 0$
and~$28$. In contrast to porosity gradients, gradients in the
strain-rate are large from the onset of simulations, as illustrated by
Figure \ref{fig:uniform_n4}c. The localized distribution of the
strain-rate variations around the inclusion presents a significant
resolution challenge for the numerical simulations. It has therefore
proven difficult to model high strains for large values
of~$\mathfrak{n}$.

\subsection{Non-uniform initial porosity}
\label{sec:non-uniform}

\begin{figure}
  \center\includegraphics[width=0.8\textwidth]
  {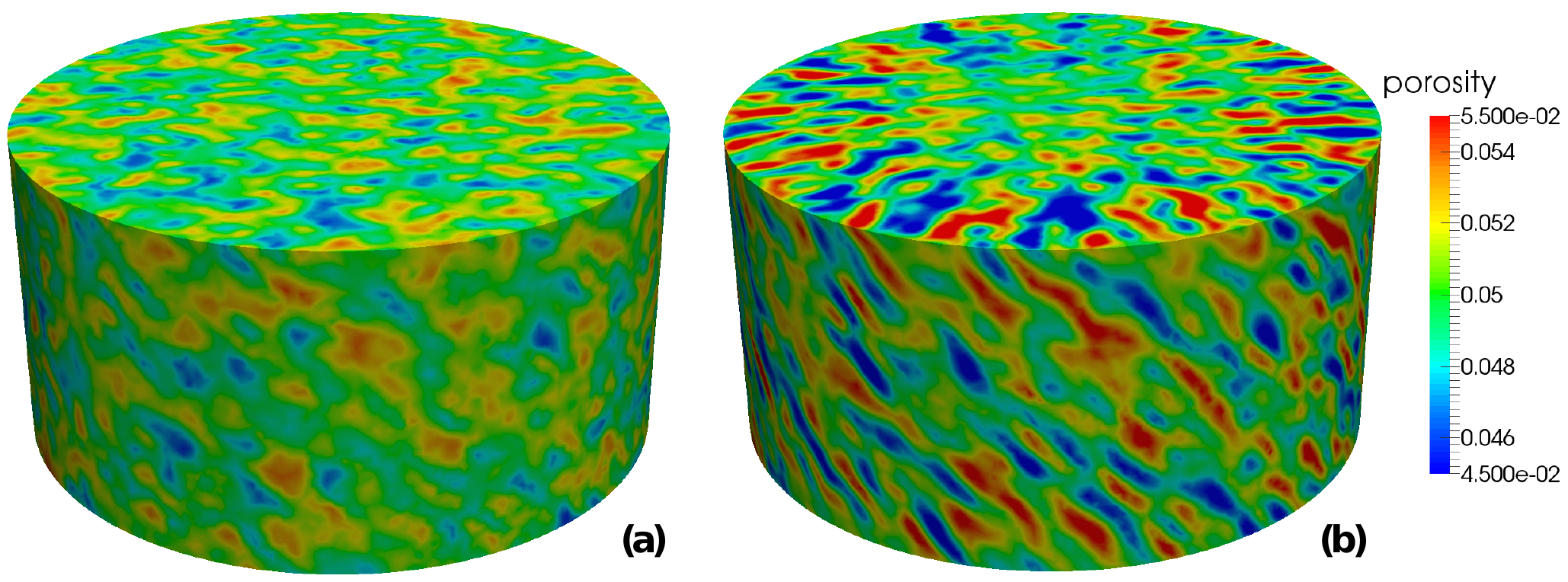}
  \caption{Example of a simulation with a random initial porosity
    field, with $\alpha$ = 28 and~$R$ = 1.7. \textbf{(a)}~Initial
    porosity field.  \textbf{(b)}~Porosity field at $t$ = 0.25.}
  \label{fig:random_cylinder}
\end{figure}

We now present simulations with a Newtonian rheology and initial
porosity perturbations with a maximum amplitude of $\pm 5 \times
10^{-3}$ about a background porosity $\phi_0$ of~$0.05$. The initial
porosity field, shown in Figure \ref{fig:random_cylinder}a, is the
same for all simulations presented in this section.

\begin{figure}
  \center\includegraphics[width=0.8\textwidth]
  {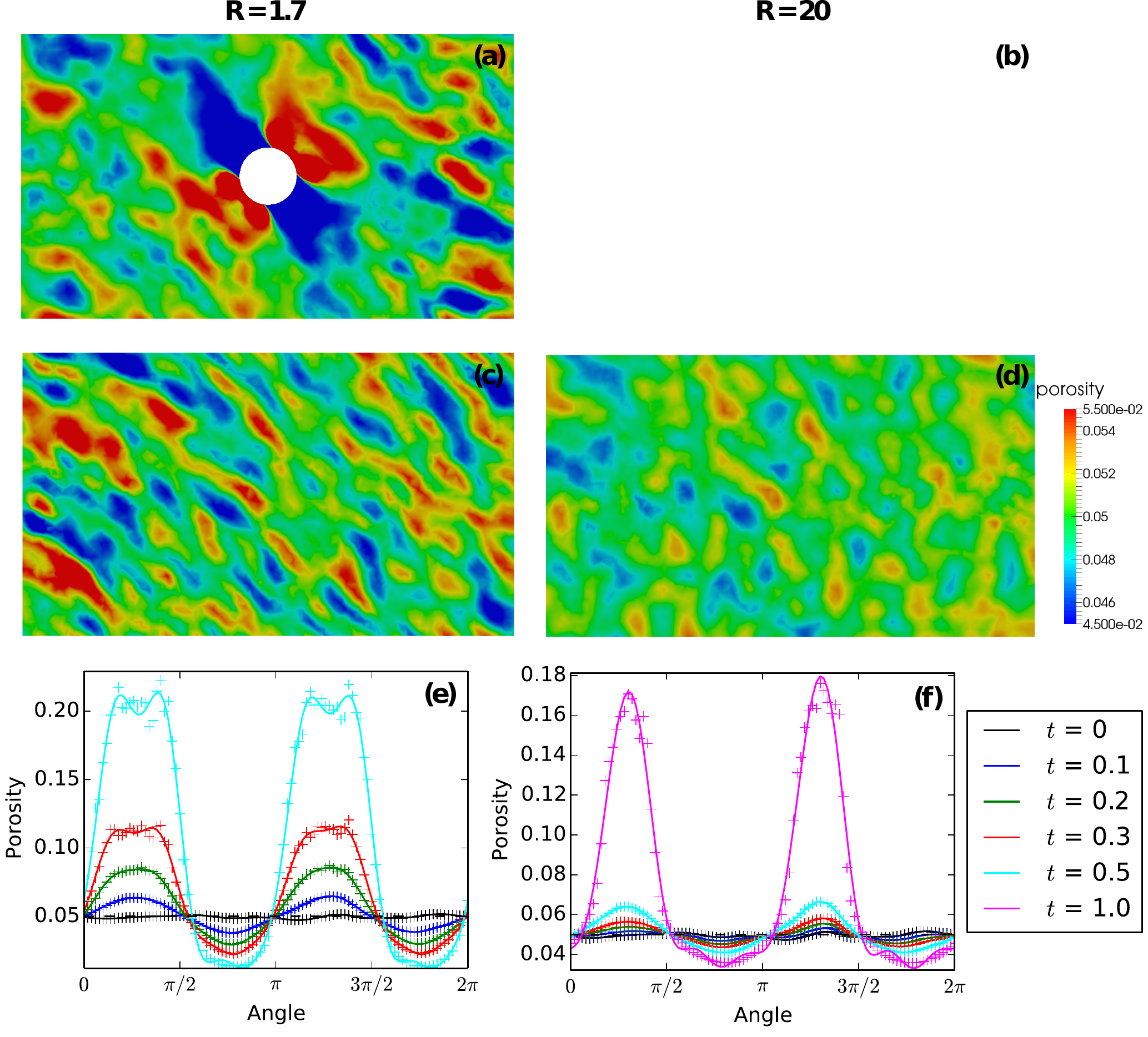}
  \caption{Results for simulations with a random initial porosity
    field and Newtonian viscosity. The local strain at the center of
    the inclusion corresponds to one half of the reported model
    time. \textbf{(a)} Slice through the porosity field on the
    inclusion side of the cylinder at $x = 1/2$, for a simulation with
    $\alpha = 28$, $R = 1.7$, at $t = 0.25$.  \textbf{(b)} Slice
    through porosity field on the inclusion side of the cylinder, for
    a simulation with $\alpha = 28$, $R = 20$, at $t = 1.0$.
    \textbf{(c)} Slice through the porosity field on the side of the
    cylinder opposite the inclusion at $x = -\frac{1}{2}$, with
    $\alpha = 28$, $R = 1.7$, at $t = 0.25$.  \textbf{(d)} Slice
    through the porosity field on the side of the cylinder opposite
    the inclusion, with $\alpha = 28$, $R = 20$, at $t = 1.0$.
    \textbf{(e)} Radial integrals over porosity, for the simulation
    with $\alpha = 28$, $R = 1.7$, at various times~$t$. The solid
    lines are fits with Fourier functions with the lowest nine
    coefficients included.  \textbf{(f)} Radial integrals over
    porosity, for the simulation with $\alpha = 28$, $R = 20$, at
    various times~$t$.}
  \label{fig:random_t}
\end{figure}

In a simulation with porosity exponent $\alpha = 28$ and
bulk-to-shear-viscosity ratio $R = 1.7$, melt-rich bands develop
throughout the cylinder over time at an angle of $\sim 45^{\circ}$
with respect to the top and bottom of the domain (Figure
\ref{fig:random_cylinder}b). Larger band amplitudes are found towards
the outside of the cylinder, as the local strain is proportional to
the radius.  Melt-rich bands develop both around the inclusion and
away from it as shown in cross-sections at $x = 1/2$ through the
inclusion and at $x = -1/2$ through the opposite side of the cylinder
in Figure \ref{fig:random_t}a and~c. In contrast, a simulation with
the same $\alpha$ and $R = 20$ does not display the formation of
melt-rich bands, as shown in Figure \ref{fig:random_t}b and~d.

The integrals in Figure \ref{fig:random_t}e--f illustrate the
difference in behavior between the $R = 1.7$ and $R=20$ simulations:
the widening and flattening of the high-porosity peaks is much more
visible in the $R = 1.7$ case than in the $R = 20$ case. In the latter
case, the porosity shadows even display an advected pattern at large
strains (represented by sharp peaks much like the simulation with a
uniform initial porosity field in Figure \ref{fig:uniform}c),
indicating that the growth of porosity is less dominant than its
advection for such large~$R$.

\begin{figure}
  \center\includegraphics[width=0.8\textwidth]
  {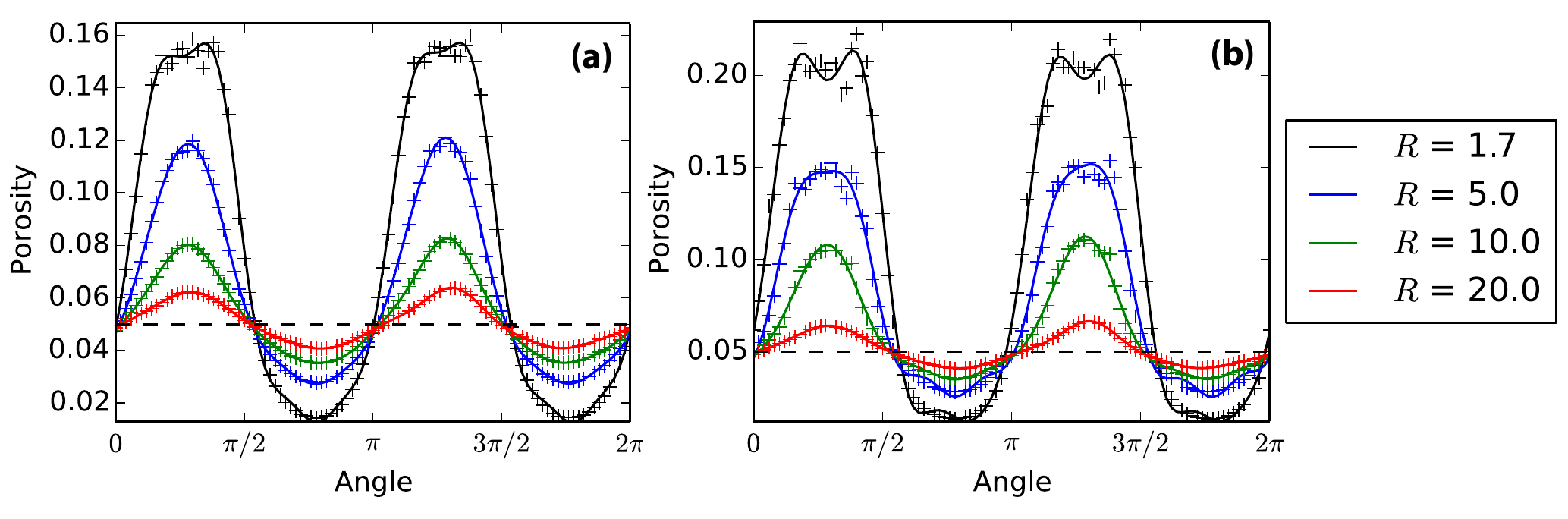}
  \caption{Radial integrals over porosity for simulations with a
    random initial porosity field and Newtonian viscosity, with
    various values of bulk-to-shear-viscosity ratio~$R$. The solid
    lines are fits with Fourier functions with the lowest nine
    coefficients included. \textbf{(a)}~Simulations with $\alpha = 15$
    at $t = 0.5$, for various values of~$R$.  \textbf{(b)}~Simulations
    with $\alpha = 28$ at $t = 0.5$, for various values of~$R$.}
  \label{fig:random_R}
\end{figure}

Melt-rich bands only develop in simulations with sufficiently large
$\alpha$ and small $R$, as illustrated by the more pronounced widening
and flattening of high-porosity peaks in the integrals in Figure
\ref{fig:random_R}. This is in line with the expected growth rates of
melt-rich bands derived using linear stability analysis and presented
in Appendix \ref{app:meltbands}. The linear stability analysis
predicts melt bands to grow initially exponentially ($\propto
\exp(\dot{s} t)$) at a dimensionless rate
\begin{equation}
    \dot{s} = \frac{\alpha \del{1 - \phi_{0}}}{R + \frac{4}{3}},
\end{equation}
which indicates that melt-rich bands are expected to grow faster for
larger $\alpha$ and smaller~$R$.

\subsection{Model regimes}
\label{sec:regimes}

\begin{figure}
  \center\includegraphics[width=0.5\textwidth]
  {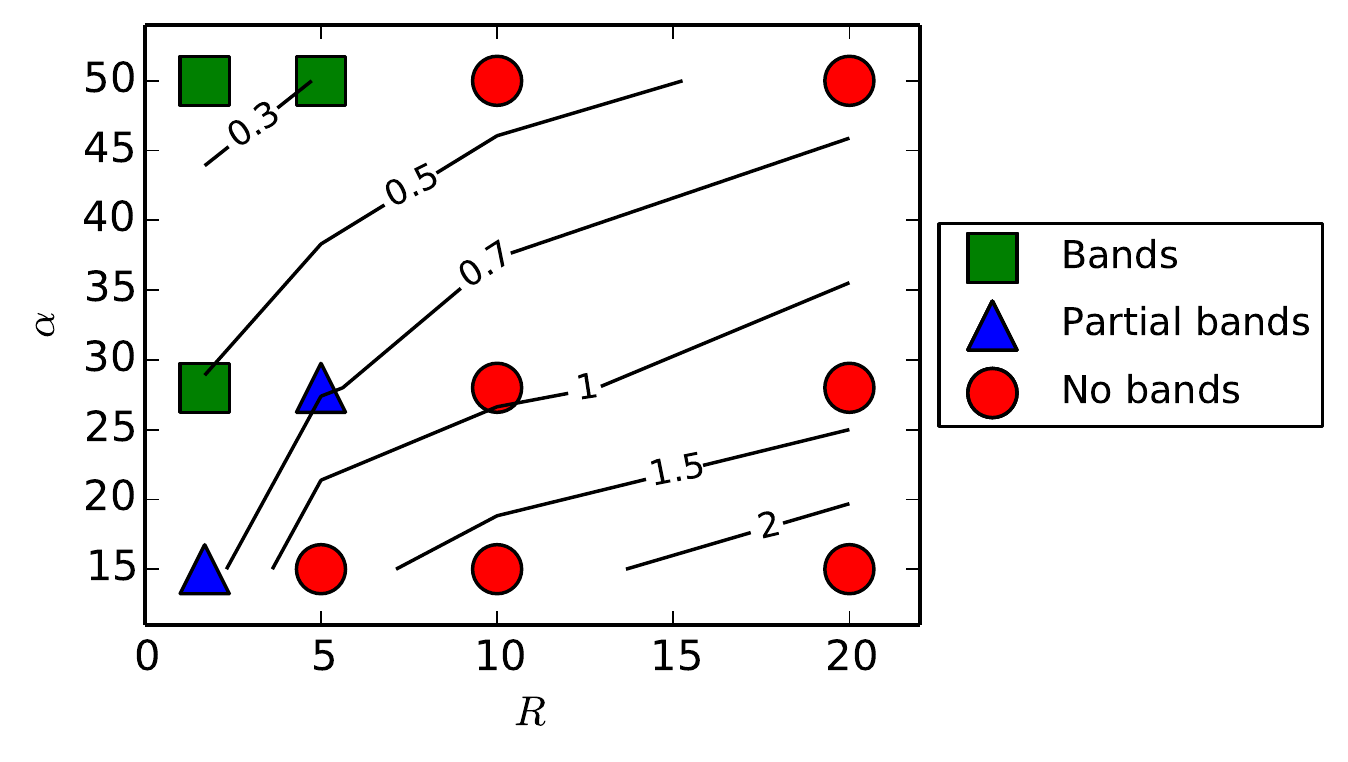}
  \caption{Summary of parameter study of bulk-to-shear-viscosity ratio
    $R$ versus porosity exponent~$\alpha$. Red circles indicate that
    no significant development of melt-rich bands takes place during a
    simulation with the specified combination of $R$ and~$\alpha$;
    green squares indicate the presence of melt-rich bands. Blue
    triangles indicate development of melt-rich bands only away from
    the inclusion. The black contours denote the maximum strain
    achieved at the outer edge of the cylinder before the porosity
    goes out of bounds and the simulations end ($\phi < 0$ or $\phi >
    1$).}
  \label{fig:regimes}
\end{figure}

Figure \ref{fig:regimes} summarizes the results of our parameter study
of porosity exponent $\alpha$ and bulk-to-shear-viscosity ratio $R$
for simulations with a random initial porosity field and Newtonian
rheology.  The overall pattern is similar to that found in the
two-dimensional study of \citet{Alisic:2014}: melt-rich bands only
develop for $R \leq 5$ and large~$\alpha$. The region of the parameter
space in which bands develop is slightly larger for the
three-dimensional geometry compared to a two-dimensional case (see
\citet[Figure~10]{Alisic:2014}), and the maximum strain achieved in
the simulations is significantly larger. This might be explained by
the fact that the amplitudes of pressure shadows decay faster away
from the inclusion in three dimensions compared to two dimensions (as
$r^{-3}$ rather than $r^{-2}$, \citet{Rudge:2014}), resulting in less
dominant pressure shadows compared to other features developing in the
porosity field.

\section{Discussion}
\label{sec:discussion}

The simulations in this study display several of the main features
observed in the laboratory experiments by \citet{Qi:2013}, such as the
pressure shadows around the spherical inclusion. For small
bulk-to-shear-viscosity ratios, melt-rich bands develop throughout the
medium, including in the vicinity of the inclusion.  However, these
bands do not have the dominance observed in the laboratory
experiments, where they grow as very straight and pervasive features
directly adjacent to the inclusion, overprinting the pressure shadows
around the inclusion. In contrast, the bands found in the numerical
simulations grow to shorter lengths and do not overprint the porosity
structure around the inclusion. Furthermore, simulations with and
without random initial porosity perturbations using a given
combination of $\alpha$ and $R$ and a Newtonian rheology run to the
same maximum model strain before going out of bounds. This behavior,
along with the fact that melt bands form for a larger $(\alpha, R)$
parameter space away from the inclusion than near it, are indications
that in our models the pressure shadows around the inclusion are
dominant over any bands that form in simulations with random
heterogeneities.  Several recent studies investigate alternative
constitutive relations \citep{Takei:2013, Katz:2013, Rudge:2015} that
could potentially affect the balance of pressure shadows and melt-rich
band formation near the inclusion.

The dominance of pressure shadows also points to a key deficiency in
current models of two-phase-flow: the models contain no physics that
sufficiently limit porosity growth, which results in the porosity
field in our models growing until reaching unity, at which time the
simulations are terminated. Realistically, the governing equations are
only valid for porosities much smaller than unity.  A second
consequence of the porosity weakening rheology in our model is the
lack of a minimum length scale (width) to which melt-rich bands will
evolve.  This means that the thickness of bands in simulations is
ultimately dictated by the grid spacing, and therefore the solutions
are resolution dependent.

It should be noted that our numerical simulations have a different
velocity boundary condition on the sides of the cylinder than the
experiments by \citet{Qi:2013}. In those experiments, velocity is only
prescribed on the top and bottom of the cylinder, and the side
boundary can slip freely. In contrast, in our simulations the velocity
is fully prescribed on all outside boundaries, leading to a
potentially more constrained model.

In simulations with a rheology that is also strain-rate-dependent,
limitations on numerical resolution near the inclusion prevented
evolution to high strains.  Therefore the regime where amplitudes of
porosity variations were large enough to allow the porosity-weakening
to become dominant over strain-rate effects was typically not reached,
and bands would not develop within the simulation time.

In future numerical studies, it would be helpful to utilize a
significantly higher resolution near the inclusion in such
simulations, so that the strain-rate gradients can be better
resolved. In addition, much could be gained from obtaining
higher-resolution images of pressure shadows in experiments: the
details of the shape of the shadows could help with identification of
the prevailing deformation mechanism (diffusion creep or dislocation
creep with large power-law exponent~$\mathfrak{n}$).

\section{Conclusions}
\label{sec:conclusions}

We have modeled the behavior of partially molten material with an
inclusion under torsion using three-dimensional numerical solutions to
the equations of two-phase flow. Recent laboratory experiments with a
similar setup display a competition between pressure shadows forming
around the inclusion and melt-rich bands that develop throughout the
partially molten medium. In our numerical simulations, the pressure
shadows around the inclusion are reproduced for all tested
combinations of bulk-to-shear-viscosity ratio and porosity exponent of
the shear viscosity. In contrast, melt-rich bands only develop for
small bulk-to-shear-viscosity ratios of five or less. We conclude that
it is more difficult to form melt-rich bands near the inclusion, which
provides a strong perturbation to the pressure field in the form of
pressure shadows.  Comparing this study with our earlier work in two
dimensions, we show that the pressure shadows are less dominant in
three dimensions, resulting in more pervasive development of melt-rich
bands. For strain-rate dependent viscosity, the shape of the pressure
shadows is significantly different compared to Newtonian cases. This
variation in shape could be utilized to pinpoint the dominant
deformation mechanism around the inclusion in future experiments.

\appendix
\section{Pressure splitting and nondimensionalization of the governing
equations}
\label{sec:nondim}

The dimensional equations for two-phase flow are:
\begin{align}
  \frac{\partial \phi}{\partial t}
  - \nabla \cdot \del{1 - \phi} \mathbf{u}_{s} &= 0,
  \label{eq:mck1}
  \\
  \nabla \cdot \Bar{\mathbf{u}} &= 0,
  \label{eq:mck2}
  \\
  \phi(\mathbf{u}_{f} - \mathbf{u}_{s})
  &= - \frac{K_{\phi}}{\mu_{f}} \nabla p_{f},
  \label{eq:mck3}
  \\
  \nabla \cdot \Bar{\boldsymbol{\sigma}} &= \mathbf{0},
  \label{eq:mck4}
\end{align}
where $\phi$ denotes porosity, $t$ time, and $\mathbf{u}_s$ and
$\mathbf{u}_f$ the solid and fluid velocities, respectively. Bulk
properties are denoted with an overbar, where a bulk quantity $\bar{a}
= \phi a_f + (1 - \phi) a_s$. Furthermore, $K_{\phi}$ is the
permeability, $\mu_f$ the fluid viscosity, $p_f$ the fluid pressure,
and $\Bar{\boldsymbol{\sigma}}$ is the bulk stress tensor.

We define the bulk stress tensor in terms of the fluid pressure $p_f$,
compaction pressure $p_c$, and the deviatoric stress
tensor~$\Bar{\boldsymbol{\tau}}$:
\begin{align}
  \Bar{\boldsymbol{\sigma}} &= - p_f \mathbf{I} - p_c \mathbf{I}
  + \Bar{\boldsymbol{\tau}},
  \\
  p_c &= - \zeta \nabla \cdot \mathbf{u}_{s},
  \\
  \Bar{\boldsymbol{\tau}} &= 2 \eta \dot{\boldsymbol{e}}
  = \eta \left( \nabla \mathbf{u}_{s} + \left(\nabla \mathbf{u}_{s}\right)^T
  - \frac{2}{3} \left( \nabla \cdot \mathbf{u}_{s} \right) \mathbf{I} \right),
\end{align}
where $\mathbf{I}$ is the identity tensor, $\zeta$ the bulk viscosity,
$\eta$ the shear viscosity, and $\dot{\boldsymbol{e}}$ the deviatoric
strain rate tensor.

We can now write a new system of equations using $\mathbf{u}_s$,
$p_f$, $p_c$, and $\phi$ as unknowns:
\begin{align}
  \frac{\partial \phi}{\partial t} -
  \nabla \cdot \del{1 - \phi} \mathbf{u}_{s} &= 0,
  \\
  - \nabla \cdot \mathbf{u}_s +
  \nabla \cdot \left( \frac{K_{\phi}}{\mu_{f}} \nabla p_{f} \right) &= 0,
  \\
  - \nabla \cdot \mathbf{u}_{s} - \zeta^{-1} p_c &= 0,
  \\
  - \nabla \cdot \Bar{\boldsymbol{\tau}}
  + \nabla p_f + \nabla p_c &= \mathbf{0}.
\end{align}
Constitutive properties are defined in this study as follows:
\begin{align}
  K_{\phi} &= K_{\mathrm{ref}} \left( \frac{\phi}{\phi_0} \right)^n,
  \\
  \eta &= \eta_{\mathrm{ref}}
  \left(\frac{\dot{{\varepsilon}}}{\dot{{\varepsilon}}_\text{ref}} \right)^{-q}
  \mathrm{e}^{-\alpha (\phi - \phi_0)},
  \label{eq:eta_law}
  \\
  \zeta &= \zeta_{\mathrm{ref}}
  \left( \frac{\dot{{\varepsilon}}}{\dot{{\varepsilon}}_\text{ref}}
  \right)^{-q}
  \mathrm{e}^{-\alpha (\phi - \phi_0)} \left( \frac{\phi}{\phi_0}
  \right)^{-m}
  = R \eta \left( \frac{\phi}{\phi_0} \right)^{-m},
  \label{eq:zeta_law}
\end{align}
where $n = 2$, $m = 1$, and $\alpha$ is the porosity exponent and
$K_{\mathrm{ref}}$ denotes the permeability at the reference
porosity~$\phi_0$.  The second invariant of the deviatoric strain-rate
tensor $\dot{{\varepsilon}}$ is given by:
\begin{equation}
  \dot{{\varepsilon}} = \left(\frac{1}{2} \dot{\boldsymbol{e}}:
  \dot{\boldsymbol{e}} \right)^{1/2},
\end{equation}
and $q$ is related to the power law exponent $\mathfrak{n}$ by
\begin{equation}
 q = 1 - \frac{1}{\mathfrak{n}}.
\end{equation}
The reference value of the second invariant is chosen as
\begin{equation}
  \dot{{\varepsilon}}_\mathrm{ref} = \frac{\dot{\gamma} \rho}{2 H}.
\end{equation}
which is the value the second invariant takes on the curved boundary
of the cylinder under the imposed torsion field. $H$ is the radius of
the cylinder and $\dot{\gamma}$ is the imposed shear strain rate on
the curved boundary. $\eta_{\mathrm{ref}}$ and $\zeta_{\mathrm{ref}}$
thus represent the shear and bulk viscosities at the curved boundary
when the porosity is uniform and equal to~$\phi_0$.  The
bulk-to-shear-viscosity ratio $R$ is given by~$\zeta_{\mathrm{ref}} /
\eta_{\mathrm{ref}}$.

To complete the problem, boundary conditions are applied as follows:
\begin{align}
  - \frac{K_{\phi}}{\mu_{f}} \nabla p_{f} \cdot \mathbf{n}
  &= 0 \ \text{on} \ \partial\Omega,
  \label{eq:bc1} \\
  \mathbf{u}_{s} &= \mathbf{w} \ \text{on} \ \partial\Omega,
  \label{eq:bc2}
\end{align}
where $\mathbf{w}$ is a prescribed solid velocity, and the boundaries
are taken to be impermeable.

We must now define a convention for nondimensionalizing the governing
equations, using primes for dimensionless quantities:
\begin{equation}
  \begin{gathered}
    \mathbf{x} = H \mathbf{x}',
    \quad
    \mathbf{u_s} = H \dot {\gamma} \mathbf{u_s}',
    \quad
    t = \dot{\gamma}^{-1} t',
    \\
    p_{f} = \eta_{\mathrm{ref}} \dot{\gamma} p_{f}',
    \quad
    p_{c} = \eta_{\mathrm{ref}} \dot{\gamma} p_{c}',
    \\
    K_{\phi} = K_{\mathrm{ref}} K_{\phi}',
    \quad
    \eta = \eta_{\mathrm{ref}} \eta',
    \quad
    \zeta = \zeta_{\mathrm{ref}} \zeta'.
  \end{gathered}
\end{equation}

The dimensionless system of equations then becomes, after dropping the
primes, equations \eqref{eq:sys1_nondim}--\eqref{eq:sys4_nondim} in
the main text.

\section{Analysis and code benchmarks}
\label{sec:analysis}

\subsection{Instantaneous compaction around an inclusion}
\label{sec:instcomp}

The instantaneous rate of compaction around a spherical inclusion in
an unbounded medium with uniform porosity can be derived analytically
\citep{McKenzie:2000,Rudge:2014}, which therefore lends itself for
benchmarking our numerical method.

In \citet{Rudge:2014} analytical solutions were presented for
compacting flow past a sphere in far-field simple shear. Here we
generalize this solution to the case of a far-field torsional flow. We
follow the approach of \citet{Rudge:2014} in this appendix, which has
a different set of coordinates to the main body of this paper, with
the center of the sphere being the origin of the coordinate system. In
Cartesian coordinates the far-field torsional flow takes the form
\begin{equation}
  \mathbf{u}_{s}^\infty = \dot{\varGamma}\del{-(y - y_0) (z - z_0), (x - x_0)
  (z - z_0), 0},
\end{equation}
where the origin of the torsional flow is at $(x_0, y_0, z_0)$ and
$\dot{\varGamma}$ is the twist rate. The above can be decomposed into
irreducible Cartesian tensors as
\begin{equation}
  \mathbf{u}_{s}^\infty = \boldsymbol{V} + \boldsymbol{\varOmega} \times
  \boldsymbol{x} +
  \boldsymbol{\mathsf{E}} \boldsymbol{\cdot} \boldsymbol{x} -\frac{1}{3}
  \boldsymbol{x} \times
  \left( \boldsymbol{\mathsf{\theta}} \boldsymbol{\cdot} \boldsymbol{x}
  \right),
\end{equation}
where
\begin{gather}
  \boldsymbol{V} = \dot{\varGamma} \left(-y_0 z_0, \, x_0 z_0, \, 0 \right),
  \\
  \boldsymbol{\varOmega} = \dot{\varGamma} \left( \dfrac{x_0}{2}, \, \dfrac{y_0}{2}, \, - z_0 \right),
  \\
  \boldsymbol{\mathsf{E}} = \dot{\varGamma} \left( \begin{array}{ccc}
    0 & 0 & y_0/2 \\
    0 & 0 & -x_0/2\\
    y_0/2 & -x_0/2 & 0
  \end{array} \right),
  \\
  \boldsymbol{\mathsf{\theta}} = \dot{\varGamma} \left( \begin{array}{ccc}
    -1 & 0 & 0 \\
    0 & -1 & 0 \\
    0 & 0 & 2
  \end{array} \right),
\end{gather}
and $\boldsymbol{x} = (x, y, z)$ is the position vector. According to
the Fax\'{e}n laws a sphere placed at the origin in such a flow will
translate with velocity $\boldsymbol{V}$ and rotate with angular
velocity $\boldsymbol{\varOmega}$, provided that there is no net force
or torque on the sphere. The compacting flow past the sphere can be
calculated as a linear superposition of the flow due to pure strain
$\boldsymbol{\mathsf{E}} \boldsymbol{\cdot} \boldsymbol{x}$
\citep[Section~5]{Rudge:2014} and that due to the vortlet flow
$-\frac{1}{3} \boldsymbol{x} \times \left(
\boldsymbol{\mathsf{\theta}} \boldsymbol{\cdot} \boldsymbol{x}
\right)$ (a quadratic flow, not considered in \citet{Rudge:2014}). The
vortlet flow is characterized by the second-rank pseudo-tensor
$\boldsymbol{\mathsf{\theta}}$ which is equal to the vorticity
gradient. The perturbation flow satisfies boundary conditions
\begin{gather}
  \tilde{\mathbf{u}}_s |_{r=a} = \frac{1}{3} \boldsymbol{x} \times \left(
  \boldsymbol{\mathsf{\theta}} \boldsymbol{\cdot} \boldsymbol{x} \right),
  \quad \quad \boldsymbol\nabla \tilde{p}_f \boldsymbol{\cdot} \mathbf{n}
  |_{r=a} = 0,
  \\
  \tilde{\mathbf{u}}_s \rightarrow \boldsymbol{0}, \quad \tilde{p}_f
  \rightarrow 0,
  \quad \text{as } r \rightarrow \infty,
\end{gather}
where $r = | \boldsymbol{x} |$ is distance from the center of the
sphere, and $a$ is the radius of the sphere.  The solution to the
governing equations with these boundary conditions does not involve
compaction and is simply a Stokes flow, given by
\begin{gather}
  \tilde{\mathbf{u}}_s = \frac{a^5}{3 r^5} \boldsymbol{x} \times \left(
  \boldsymbol{\mathsf{\theta}} \boldsymbol{\cdot} \boldsymbol{x} \right),
  \\
  \tilde{p}_f =0.
\end{gather}

Since the quadratic flow does not involve compaction, the
instantaneous compaction rate for a sphere in a torsional field is the
same as that for pure and simple shear, given by \citet[eqns.~(5.51)
  and (5.63)]{Rudge:2014}. When the compaction length is large
compared to the domain size, the behavior can be well described by the
large-compaction-length asymptotic limit of the equations, where the
instantaneous compaction rate and fluid pressure are given by
\begin{gather}
  \nabla \cdot \mathbf{u}_{s}
  = \frac{15\nu}{2\nu + 3} \del{\frac{a}{r}}^3 \frac{\boldsymbol{x}
  \boldsymbol{\cdot} \boldsymbol{\mathsf{E}} \boldsymbol{\cdot}
  \boldsymbol{x}}{r^2},
  \label{eq:an_crate}
  \\
  p_f = \frac{\mu_f a^2}{K_{\mathrm{ref}}}
  \frac{5 \nu}{6 \del{2\nu + 3}} \left[ \del{\frac{a}{r}}^3
    - \frac{3 a}{r} \right] \frac{\boldsymbol{x} \boldsymbol{\cdot}
    \boldsymbol{\mathsf{E}} \boldsymbol{\cdot} \boldsymbol{x}}{r^2},
  \label{eq:an_pressure}
\end{gather}
where $\nu \equiv \eta_\mathrm{ref} / \del{\zeta_\mathrm{ref} +
  4\eta_\mathrm{ref}/3} = \del{R + 4/3}^{-1}$. The above solution is
identical to that given in equations~(30) and~(32) of
\citet{McKenzie:2000}.

\begin{figure}
  \center\includegraphics[width=0.8\textwidth]
  {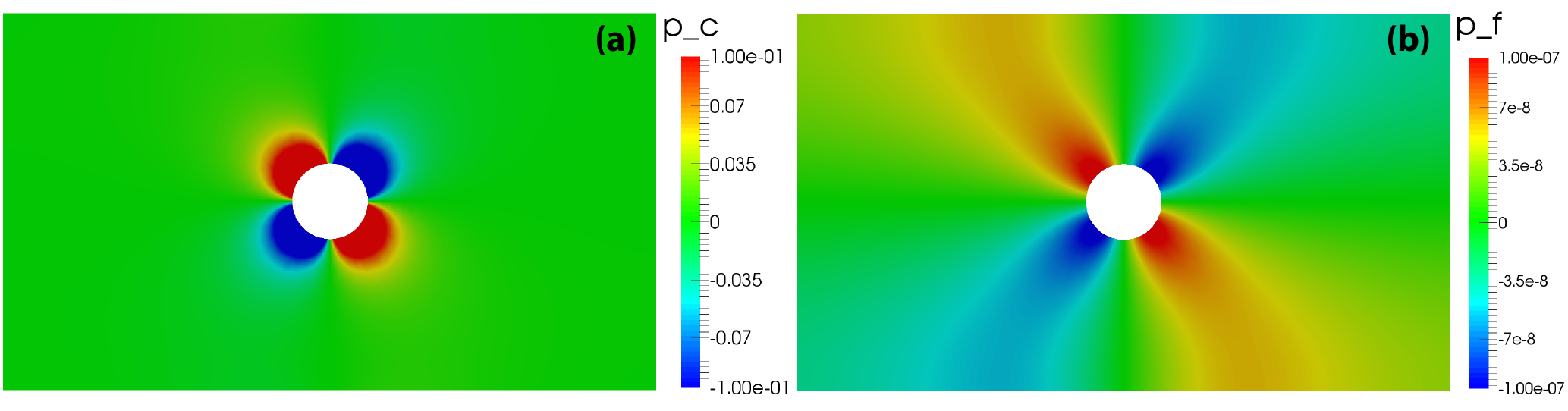}
  \caption{Instantaneous pressure fields for a simulation with
    inclusion size~$0.1$, porosity exponent $\alpha$ = 28,
    bulk-to-shear-viscosity ratio $R$ = 5/3, and stress exponent
    $\mathfrak{n}$ = 1.  \textbf {(a)}~Compaction
    pressure. \textbf{(b)}~Fluid pressure.}
  \label{fig:compaction_slicecontours}
\end{figure}

In Figure \ref{fig:compaction_slicecontours} we show the
numerically-calculated instantaneous compaction pressure and fluid
pressure on the two-dimensional slice indicated in Figure
\ref{fig:3Dgeometry}b, resulting from the imposed torsion. We use a
large compaction length with $D$ = 100, and the same cylindrical mesh
as outlined in Section \ref{sec:domain}. The numerically-calculated
compaction pressure, shown in Figure
\ref{fig:compaction_slicecontours}a, very closely matches the
analytical expression in~\eqref{eq:an_crate}. The fluid pressure,
shown in Figure \ref{fig:compaction_slicecontours}b, matches less
well, with the expected quadrupole pattern of \eqref{eq:an_pressure}
disturbed at the top and bottom boundaries of the finite computational
domain. That the fluid pressure is affected more by the boundaries
than the compaction pressure is to be expected from the analytical
expressions in \eqref{eq:an_crate} and~\eqref{eq:an_pressure};
compaction pressure decays rapidly away from the inclusion, as
$r^{-3}$, whereas fluid pressure decays much more slowly, as $r^{-1}$,
and thus the fluid pressure feels effects at larger distances.

\begin{figure}
  \center\includegraphics[width=0.5\textwidth]
  {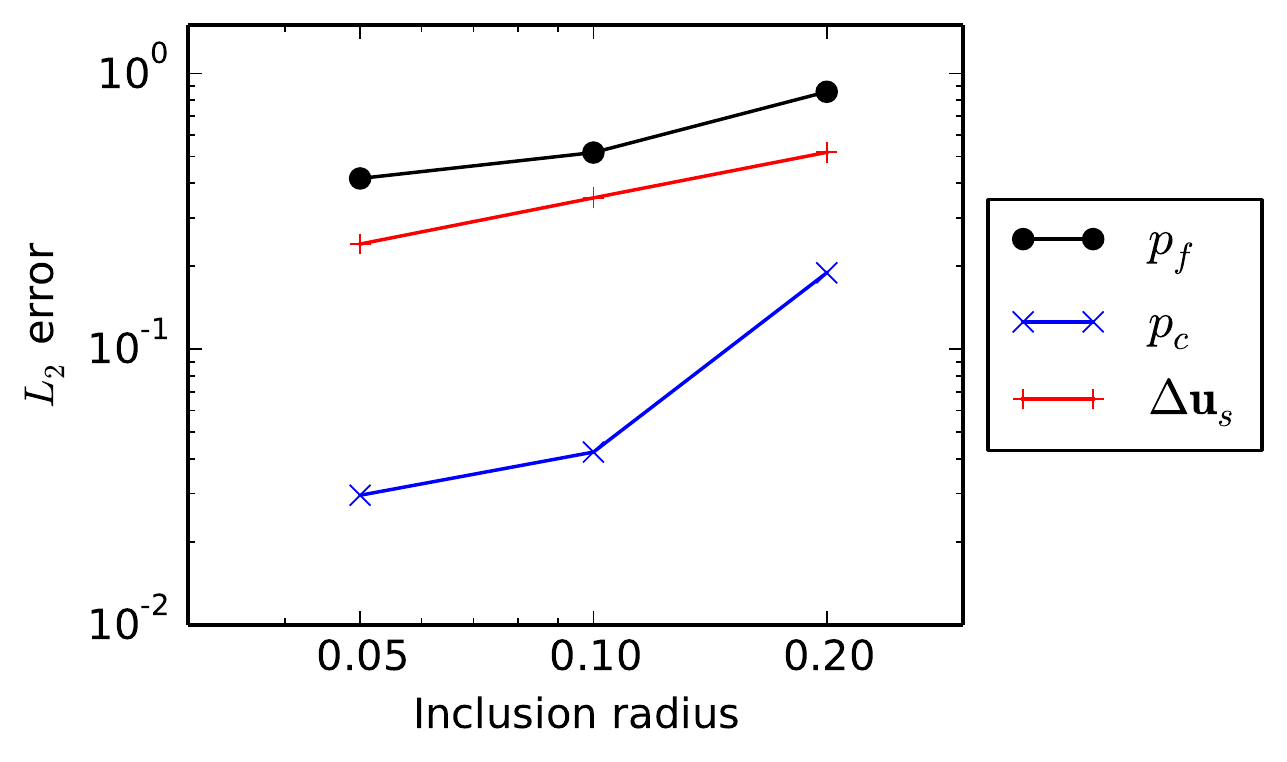}
  \caption{$L_2$ error norms computed for simulations with different
    inclusion radii, for fluid pressure $p_f$, compaction pressure
    $p_c$, and solid velocity perturbation $\Delta \mathbf{u}_s$ with
    respect to a torsional velocity field without an inclusion.}
\label{fig:compaction_error_norms}
\end{figure}

For further validation of the numerical simulations, we compute $L_2$
error norms for the fluid pressure $p_f$, compaction pressure $p_c$,
and solid velocity $\mathbf{u}_s$, with respect to the analytical
solutions. We define the following error for a field $\chi$:
\begin{equation}
  e_{L_2} = \frac{|| \chi^N - \chi^A ||_2} {||\chi^A||},
  \label{eq:L2_norm}
\end{equation}
where the numerical field is denoted by $\chi^N$ and the analytical
solution by $\chi^A$. We compute this for a series of inclusion radii
between $0.05$ and $0.2$, as shown in Figure
\ref{fig:compaction_error_norms}. The $L_2$ error decreases with
decreasing inclusion radius. This indicates that the error with
respect to the analytical solution results from the presence of
boundaries in the numerical domain, which do not exist in the
analytical solution. This effect becomes less dominant when the
inclusion is further away from the outside cylinder boundaries, i.e.,
for smaller inclusions. In addition, the $L_2$ error is larger for the
fluid pressure than for the compaction pressure. This is due to the
same boundary effects as observed in Figure
\ref{fig:compaction_slicecontours}.

\subsection{Linear stability analysis of melt bands under torsion}
\label{app:meltbands}

Linear stability analysis provides important insight into the expected
growth rate of melt bands \citep[e.g.][]{Stevenson:1989,
  Spiegelman:2003, Katz:2006, Butler:2009, Takei:2013,
  Rudge:2015}. The solutions arising from linear stability analysis
can also be used as a check on numerical solutions of the full set of
governing equations (see \citet[Appendix~C2]{Alisic:2014}). In this
appendix we present a linear stability analysis of melt bands under
torsion in an infinite cylinder for our chosen rheology, and use the
solutions to benchmark our numerical code. The analysis closely
follows an earlier linear stability analysis of melt bands under
torsion by \citet{Takei:2013}. In \citet{Takei:2013} it was assumed
that perturbation wavenumbers are large, which allows the neglect of
various radial derivatives in the analysis. We do not make this
assumption here, and instead solve numerically for the radial
variation.

The base state solution has a uniform porosity $\phi_0$ and solid
velocity $\mathbf{u}_{0}$ given in cylindrical coordinates $(\rho,
\psi, z)$ as
\begin{equation}
  \mathbf{u}_{0} =\dot{\varGamma} \rho z \hat{\boldsymbol{\psi}},
\end{equation}
where $\dot{\varGamma}$ is the twist rate. The twist rate
$\dot{\varGamma}$ is related to the shear strain rate $\dot{\gamma}$
on the cylinder edge by $\dot{\gamma} = \dot{\varGamma} H$. The base
state solution has zero pressure everywhere ($p_0 = 0$), is not
compacting such that $\mathcal{C}_0\equiv\boldsymbol\nabla \cdot
\mathbf{u}_{0}=0$, and has a strain-rate-tensor with only the $(\psi,
z)$ component non-zero,
\begin{equation}
  \dot{\boldsymbol{e}}_0
  = \dot{\varGamma} \frac{\rho}{2} \del{\hat{\boldsymbol{\psi}}
  \hat{\boldsymbol{z}}
    + \hat{\boldsymbol{z}} \hat{\boldsymbol{\psi}}}.
\end{equation}

We seek small perturbations about this base state of the form
\begin{align}
  \phi &= \phi_0 + \epsilon \phi_1 + \cdots,
  \label{eq:lin1}
  \\
  \mathbf{u}_{s} &= \mathbf{u}_0 + \epsilon \mathbf{u}_1 + \cdots,
  \label{eq:lin2}
  \\
  p_f &= 0 + \epsilon p_1 + \cdots.
  \label{eq:lin3}
\end{align}

Substituting \eqref{eq:lin1}--\eqref{eq:lin3} into the governing
equations \eqref{eq:mck1}--\eqref{eq:mck4} leads to equations at first
order in $\epsilon$ given by
\begin{gather}
  \frac{D_0 \phi_1}{Dt} = (1 - \phi_0) \mathcal{C}_1,
  \label{eq:dphidt1}
  \\
  \mathcal{C}_1 - \frac{K_0}{\mu_f} \nabla^2 p_1 = 0,
  \label{eq:eig1}
  \\
  \boldsymbol\nabla \cdot \Bar{\boldsymbol{\sigma}}_1 = \boldsymbol{0},
  \label{eq:eig2}
  \\
  \Bar{\boldsymbol{\sigma}}_1 = -p_1 \mathbf{I}
  + \zeta_0 \mathcal{C}_1 \mathbf{I} + 2 \eta_0 \dot{\boldsymbol{e}}_1
  + 2 \eta_1 \dot{\boldsymbol{e}}_0,
  \label{eq:eig3}
\end{gather}
where $\frac{D_0}{Dt} \equiv \frac{\partial}{\partial t} +
\mathbf{u}_{0} \cdot \boldsymbol\nabla$.

When the rheology is non-Newtonian, the base state viscosities vary
with radius according to
\begin{align}
  \eta_0 &= \eta_\text{ref} \left( \rho / H \right)^{-q},
  \\
  \zeta_0 &= \zeta_\text{ref} \left( \rho / H\right)^{-q}.
\end{align}
Expanding the rheological law \eqref{eq:eta_law} to first order in
$\epsilon$ yields
\begin{equation}
  \eta_1
  = \eta_0 \del{-\alpha\phi_1 - q \frac{\dot{\boldsymbol{e}}_1 :
  \dot{\boldsymbol{e}}_0}{2 \dot{{\varepsilon}}_0^2}}
  =  \eta_0 \del{-\alpha \phi_1
  - q \frac{\dot{e}_{1 \psi z}}{\dot{{\varepsilon}}_0}}.
  \label{eq:eig4}
\end{equation}

The expected growth rate of bands can be determined by replacing
\eqref{eq:dphidt1} by
\begin{equation}
  \dot{s} \phi_1 = (1 - \phi_0) \mathcal{C}_1,
  \label{eq:eig5}
\end{equation}
and solving the eigenvalue problem described by
\eqref{eq:eig1}--\eqref{eq:eig3} and \eqref{eq:eig4}--\eqref{eq:eig5}
for the instantaneous growth rate $\dot{s}$. The finite element method
can be used to numerically solve for the eigenfunctions and
eigenvalues of the linear stability equations. Equations
\eqref{eq:eig1}--\eqref{eq:eig3} and \eqref{eq:eig4}--\eqref{eq:eig5}
can be cast into a weak form for trial functions $(\boldsymbol{u}, p)$
and test functions $(\boldsymbol{v}, q)$ as
\begin{gather}
  \int_{V} 2 \eta_0 \dot{\boldsymbol{e}}^u : \dot{\boldsymbol{e}}^v +
  \zeta_0 \mathcal{C}^u \mathcal{C}^v
  - p \mathcal{C}^v - q  \mathcal{C}^u
  - \frac{K_0}{\mu_f} \boldsymbol\nabla
  p \cdot \boldsymbol\nabla q
  - 4 \eta_0 q \dot{e}_{\psi z}^u \dot{e}_{\psi z}^v \dif V
  = \lambda \int_{V} 4 \eta_0 \dot{{\varepsilon}}_0 \mathcal{C}^u
  \dot{e}_{\psi z}^v  \dif V
  \label{eq:linstabweakform},
  \\
  \dot{\boldsymbol{e}}^u \equiv \tfrac{1}{2} \del{\boldsymbol\nabla \mathbf{u} + \del{\boldsymbol\nabla \mathbf{u}}^T}
      - \tfrac{1}{3} \del{\boldsymbol\nabla \cdot \mathbf{u}} \mathbf{I},
  \\
  \mathcal{C}^u = \boldsymbol\nabla \cdot \mathbf{u},
\end{gather}
where the subscripts $1$ referring to the first order state have been
neglected for clarity. The eigenvalue $\lambda$ is related to the
growth rate $\dot{s}$ by
\begin{equation}
 \lambda = \frac{\alpha (1-\phi_0)}{\dot{s}}.
\end{equation}
The impermeable and no-slip boundary conditions \eqref{eq:bc1} and
\eqref{eq:bc2} on the cylinder edge lead to the vanishing of surface
integral terms in the weak form~\eqref{eq:linstabweakform}.

\begin{figure}
  \center\includegraphics[width=0.5\textwidth]
  {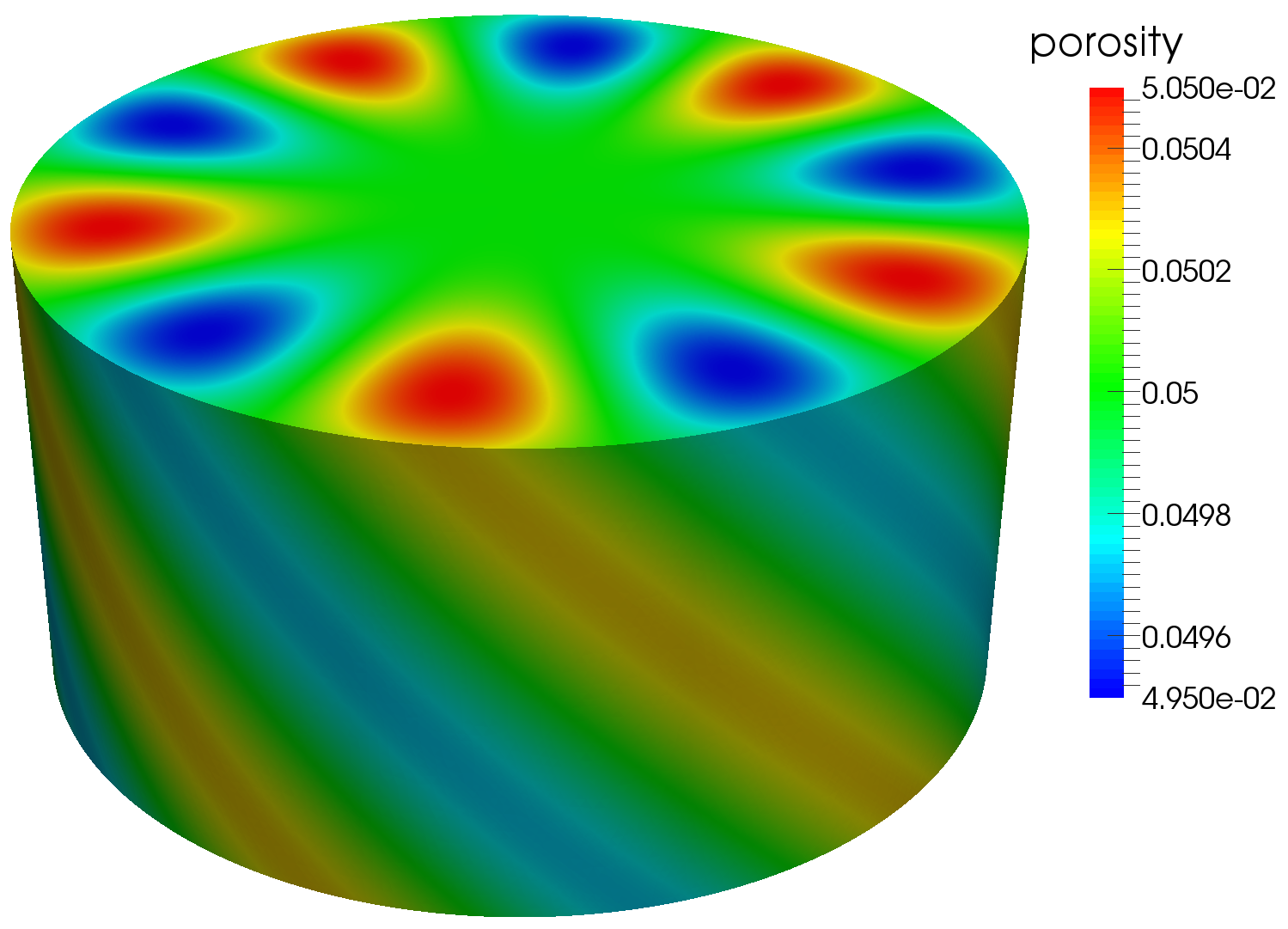}
  \caption{Porosity field for a particular eigenfunction of the linear
    stability equations. This eigenfunction has angular wavenumber $n
    = 5$, and vertical wavenumber $h$ chosen such that the angle of
    the bands to the shear plane is $45^\circ$ on the edge of the
    cylinder. The bulk-to-shear-viscosity ratio $R = 5/3$ and the
    power law exponent~$\mathfrak{n} = 1$. The compaction length is
    large (100 times the cylinder radius). The eigenfunction shown is
    the fastest growing mode with this choice of $n$ and $h$.}
\label{fig:spiral}
\end{figure}

The three-dimensional weak form \eqref{eq:linstabweakform} can be
reduced to a weak form for the radial coordinate alone using symmetry
considerations. Invariance of the cylinder under rotation about its
axis, and invariance under translation in the $z$ direction, suggests
looking for eigenfunctions proportional to $\mathrm{e}^{i n \psi + i h
  z}$ where $n$ is the angular wavenumber, and $h$ is the vertical
wavenumber. Substituting solutions of this form
\begin{align}
  u_\rho &= U_\rho(\rho) \mathrm{e}^{i n \psi + i h z},
  \\
  u_\psi &= i U_\psi(\rho) \mathrm{e}^{i n \psi + i h z},
  \\
  u_z &= i U_z(\rho) \mathrm{e}^{i n \psi + i h z},
  \\
  p &= P(\rho) \mathrm{e}^{i n \psi + i h z},
\end{align}
and corresponding complex conjugates for test functions into
\eqref{eq:linstabweakform} leads to purely radial integrals where
$\dif V \rightarrow 2 \pi \rho \dif \rho$. The integrands are real,
and the resulting radial eigenfunction problem was solved using
FEniCS/DOLFIN \citep{logg:2010} and the eigenvalue solver SLEPc
\citep{Hernandez:2005}. An example eigenfunction calculated using this
approach is shown in Figure \ref{fig:spiral}.

\subsubsection{A special case: Newtonian melt bands under torsion in an
  infinite domain}

There is a special case of the linear stability analysis for which a
complete analytical solution can be obtained. This is the case when
the rheology is Newtonian ($\mathfrak{n} = 1$), and the domain of
interest is infinite. In this case the solenoidal component of the
flow is decoupled from the irrotational component, simplifying the
analysis \citep{Spiegelman:1993, Spiegelman:2003}. The linear
stability equations are
\begin{gather}
  \frac{\partial \phi_1}{\partial t} + \dot{\varGamma} z \frac{\partial \phi_1}{\partial \psi} = (1-\phi_0)\mathcal{C}_1,
  \label{eq:phit}
  \\
  -\nabla^2 \mathcal{C}_1 + \delta^{-2} \mathcal{C}_1
  =  -2\nu\alpha \dot{\varGamma} \frac{\partial^2 \phi_1}{\partial z \partial
  \psi},
  \label{eq:c}
\end{gather}
where \eqref{eq:phit} follows from \eqref{eq:dphidt1}, and
\eqref{eq:c} is a result of combining the divergence of
\eqref{eq:eig2} with \eqref{eq:eig1}, \eqref{eq:eig3},
and~\eqref{eq:eig4}, $\delta$ is the compaction length
\eqref{eq:compaction_length}, and $\nu = \eta_\mathrm{ref} /
\del{\zeta_\mathrm{ref} + 4\eta_\mathrm{ref}/3}$. Solutions to
\eqref{eq:phit} and \eqref{eq:c} can be found in the form of
cylindrical harmonics (eigenfunctions of the Laplacian operator in
cylindricals),
\begin{gather}
  \phi_1 = \Phi(t) J_{n}(\lambda \rho) \mathrm{e}^{i n \psi + i h(t) z},
  \label{eq:phiguess}
  \\
  \mathcal{C}_1 = C(t) J_{n} (\lambda \rho) \mathrm{e}^{i n \psi + i h(t) z},
  \label{eq:cguess}
\end{gather}
where $J_{n}(z)$ is a Bessel function of the first kind and $h(t)$ is
the vertical wavenumber, which varies with time as
\begin{equation}
  h(t) = h_0 - \dot{\varGamma} n t,
\end{equation}
due to the advection. Note that
\begin{gather}
  \nabla^2 \mathcal{C}_1 = - k^2(t) \mathcal{C}_1,
  \\
  k^2(t) = \lambda^2 + h^2(t).
\end{gather}
Substituting \eqref{eq:phiguess} and \eqref{eq:cguess} into
\eqref{eq:phit} and \eqref{eq:c} yields
\begin{gather}
  \dot{\Phi}(t) = (1 - \phi_0) C(t),
  \\
  \del{\delta^{-2} + k^2(t)} C(t) = 2 \nu \alpha \dot{\varGamma} n h(t)
  \Phi(t),
\end{gather}
which can be combined to give
\begin{equation}
  \dot{\Phi} (t) = \frac{2 \nu \alpha \dot{\varGamma}
    (1-\phi_0) n h(t)}{\delta^{-2} + k^2 (t)} \Phi(t),
  \label{eq:Phievo}
\end{equation}
and integrated to give
\begin{equation}
  \Phi(t) = \del{\frac{\delta^{-2} + k^2(0)}{\delta^{-2} +
      k^2(t)}}^{\nu \alpha (1 - \phi_0)}.
  \label{eq:Phit}
\end{equation}
These expressions closely mirror the expressions for simple shear
given by \citet{Spiegelman:2003}: compare \eqref{eq:Phievo} and
\eqref{eq:Phit} here with equations~(27) and~(33), respectively, from
\citet{Spiegelman:2003}. The expressions are identical, except with
the appropriate switch of angular wavenumbers for planar wavenumbers,
and the difference in nondimensionalization.

In a cylinder of finite radius, common choices of boundary conditions
on the cylinder edge (including no-slip) lead to a coupling of the
solenoidal component of the flow and the irrotational component, which
complicates the analysis given above. Nevertheless, this special
solution can be used as a check on the numerical approach to
calculating the eigenfunctions described in the preceding section. The
special solution \eqref{eq:phiguess} and \eqref{eq:Phievo} was
recovered numerically for a choice of boundary conditions that
decouples the solenoidal flow from the irrotational. This choice of
boundary condition is impermeable, and almost, but not quite,
free-slip: on the cylinder edge $u_\rho = 0$, $\overline{\sigma}_{\rho
  z} = 0$, $\overline{\sigma}_{\rho \psi} = - {2 \nu u_\psi}/{\rho}$,
and~$\partial p_f / \partial \rho = 0$. These boundary conditions
imply that $\partial \mathcal{C} / \partial \rho = 0$ on the cylinder
edge, and restrict the allowable values of $\lambda$ to the roots of
the derivative of the Bessel function. As is the case for simple
shear, the fastest growing modes occur for infinite wavenumbers. The
maximum growth rate occurs as $n \rightarrow \infty$, $h \sim n / H$
(i.e., bands at $45^\circ$ to the shear plane on the cylinder edge)
where~$\dot{s}\rightarrow \nu \alpha (1- \phi_0) \dot{\varGamma} H$.

\subsubsection{Linear stability benchmark}

We test the application code by numerically computing the
instantaneous growth rate of porosity for an initial porosity field
given by the eigenfunction shown in Figure \ref{fig:spiral}. The
numerically computed growth rate can be compared with the expected
growth rate of the eigenfunction determined from the eigenvalue.

\begin{figure}
  \center\includegraphics[width=0.5\textwidth]
  {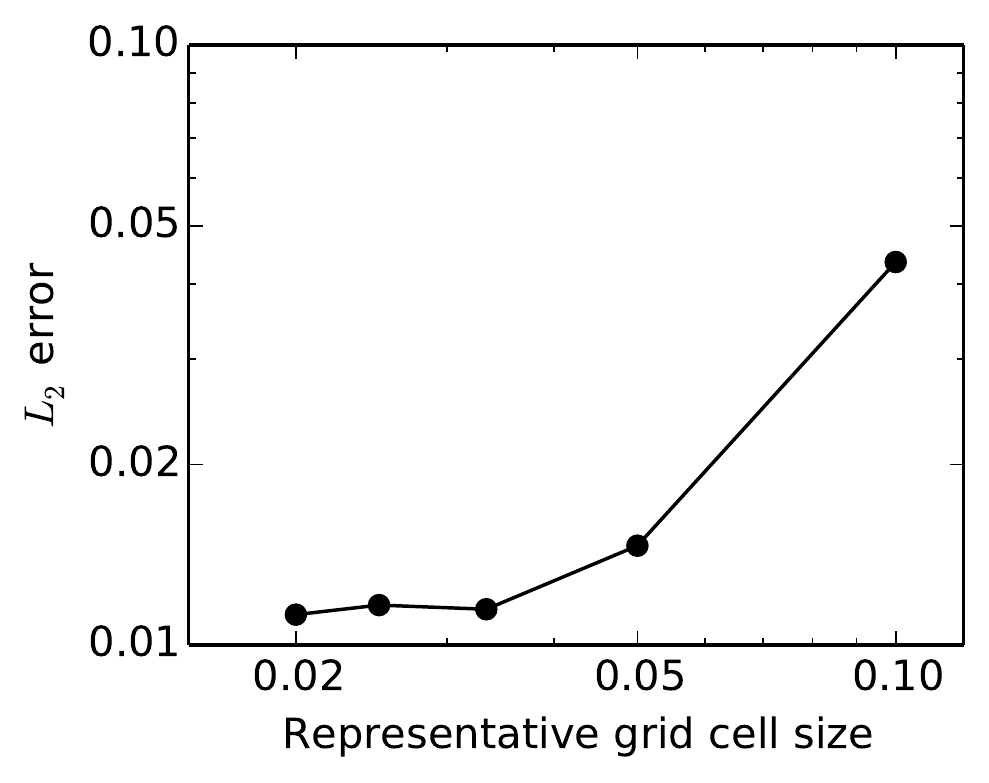}
  \caption{$L_2$ error norms of growth rate at the center-plane of the
    cylinder, computed for numerical simulations with different
    resolutions. The eigenfunction tested is shown in Figure
    \ref{fig:spiral}.}
  \label{fig:spiral_error_norms}
\end{figure}

Figure \ref{fig:spiral_error_norms} shows an example of such a
comparison, where an error norm for growth rate is plotted against
resolution for various mesh resolutions from approximately $10$ to
$50$ elements in the vertical direction, to study the effect of grid
size on the accuracy of the numerical method.
It is important to note
that the eigenfunctions are determined for a cylinder of infinite
extent, whereas the simulation domain is a cylinder of finite
extent. To mitigate the resulting boundary effects, the two are
compared only on a slice through the center-plane of the cylinder,
at~$z = 1/2$. The $L_2$ error norm is calculated for the local
instantaneous growth rate of porosity on the slice, using
equation~\eqref{eq:L2_norm}. Generally, the error in growth rate on
the center-plane decreases with increasing resolution (i.e., with
decreasing grid size), until a limit is reached at a grid size around
0.03. For finer grids the error does not decrease any further, which
we attribute to the effect of the top and bottom boundaries on the
growth of porosity. Computed growth rates are typically within a few
per cent of the expected growth rates, which gives us confidence that
the application code is solving the compaction equations effectively.

\subsection*{Acknowledgements}

This work was supported by the UK Natural Environment Research Council
under grants NE/I023929/1 and NE/I026995/1. Computations were
performed on the ARCHER UK National Supercomputing Service
(\texttt{http://www.archer.ac.uk}). We thank Chris Richardson for all
his support with running the simulations on ARCHER. Katz thanks the
Leverhulme Trust for support.


\end{document}